\documentclass[aps,pre,final,twocolumn,singlespace,groupedaddress ,nofootinbib]{revtex4-1}

\usepackage{amsthm}
\usepackage{subfigure}
\usepackage{graphicx}
\usepackage{epsfig}
\usepackage{epstopdf}
\usepackage{color}
\usepackage{natbib}
\usepackage{amsmath}
\usepackage{amssymb}
\usepackage{appendix}
\usepackage{hyperref}
\usepackage{longtable}
\usepackage{soul}
\DeclareMathOperator*{\argmax}{\arg\max}   

\begin{document}

\title{Confinement effects in premelting dynamics}


\author{Satyajit Pramanik}
\email[]{satyajit.math16@gmail.com}
\affiliation{Nordita, Royal Institute of Technology and Stockholm University, Stockholm, Sweden}

\author{John S. Wettlaufer}
\email[]{john.wettlaufer@yale.edu}
\affiliation{Yale University, New Haven, USA}
\affiliation{Mathematical Institute, University of Oxford, Oxford, UK}
\affiliation{Nordita, Royal Institute of Technology and Stockholm University, Stockholm, Sweden}


\date{\today}

\begin{abstract}
We examine the effects of confinement on the dynamics of premelted films driven by thermomolecular pressure gradients.   Our approach is to modify a well-studied setting in which the thermomolecular pressure gradient is driven by a temperature gradient parallel to an interfacially premelted elastic wall.  The modification treats the increase in viscosity associated with the thinning of films, studied in a wide variety of materials, using a power law and we examine the consequent evolution of the confining elastic wall.  We treat (1) a range of interactions that are known to underlie interfacial premelting and (2) a constant temperature gradient wherein the thermomolecular pressure gradient is a constant. The difference between the cases with and without the proximity effect arises in the volume flux of premelted liquid.  The proximity effect increases the viscosity as the film thickness decreases thereby requiring the thermomolecular pressure driven flux to be accommodated at higher temperatures where the premelted film thickness is the largest.  Implications for experiment and observations of frost heave are discussed.   
\end{abstract}

\pacs{64.70.Dv, 68.15.+e, 68.45.Gd, 82.65.Dp}
\keywords{}

\maketitle

\section{Introduction \label{sec:Intro}}

There are a host of phenomena in which the equilibrium domain of one phase is extended into the bulk region of a neighboring phase. In general, such an extension is associated with effective surface fields--forces that are extensive with area even when their underlying origins are volume-volume interactions.  Examples include, among many others, van der Waals and screened Coulomb interactions or steric effects \cite[e.g.,][]{SchickLesHouches, Safranbook, French:2010, Jacobbook}.  Here we focus on {\em premelting}, wherein a solid--wall or solid--vapor interface is wet by the melt phase of the solid in equilibrium due to the 
relative polarizabilities of a solid/liquid/wall ({\em interfacial premelting}) or solid/liquid/vapor ({\em surface melting}) interface.  One prepares the ``dry'' solid/wall or solid/vapor interface at a pressure and temperature below the bulk melting transition and approaching the latter from below.  If the interface is wet by the melt phase then, depending on the detailed nature of the underlying interactions, the film will become arbitrarily thick as the bulk transition is approached. 

\begin{figure}[htbp]
\centering 
\includegraphics[scale=0.35]{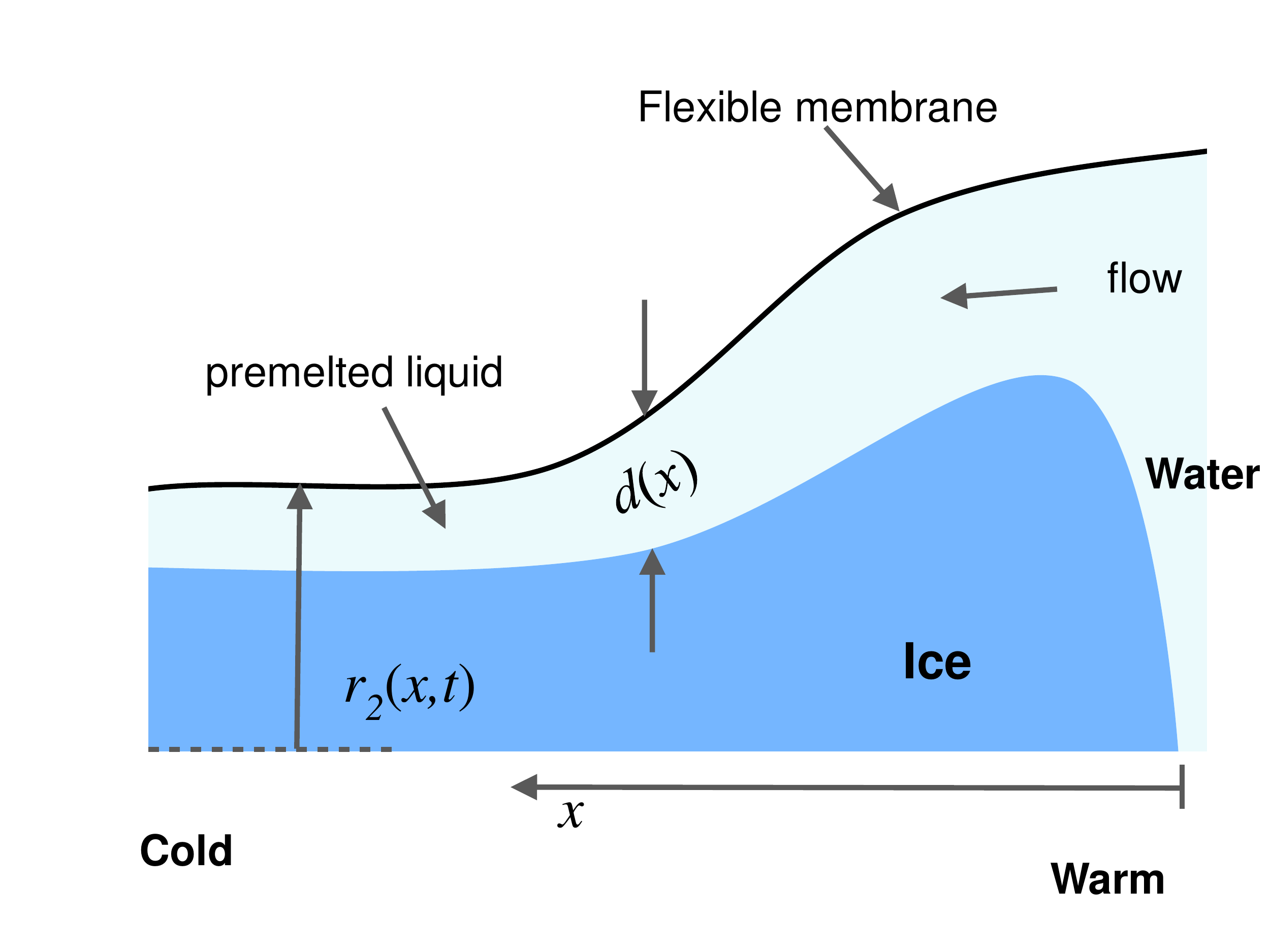} 
\caption{Schematic of the flow of a premelted film disjoining an elastic capillary tube from the solid phase from \cite{Wettlaufer2006}.} 
\label{fig:schematic}
\end{figure}

Here we focus upon interfacial premelting, which, as in the case of surface melting, occurs on at least one facet of most materials \cite{Dash2006} and has been thoroughly studied for a wide class of wall materials (conductors, dielectrics and polymers) in the case of ice \cite{Wilenetal1995}.  The liquid film ($l$) has a thickness $d$ and disjoins the solid phase ($s$) from a wall at temperatures below the bulk freezing point, $T_m$. The mean field description of the system has a grand potential 

\begin{equation}
\label{eq:free_energy}
\Omega = -P_l V_l - P_s V_s + \mathcal{I}(d). 
\end{equation}
Here $P$ and $V$ denote pressure and volume, and the volume-volume interactions are captured by the part of free energy $\mathcal{I}(d) = [\Delta \sigma f(d) + \sigma_{sw}]A_i$ that is extensive with interfacial area $A_i$, where $\Delta \sigma = \sigma_{sl} + \sigma_{lw} - \sigma_{sw}$, and the $\sigma$'s are the solid-liquid ($sl$), liquid-wall ($lw$), and solid-wall ($sw$) interfacial free energies \cite{Wettlaufer1995}. Thus, when the dry interfacial energy, $\sigma_{sw}$, is larger than $ \sigma_{sl} + \sigma_{lw}$, then sufficient conditions for the interface to be wet by the melt are met.  Power law interaction potentials have the form $f(d) = 1-(d_0/d)^{- n+1}$, where $d_0$ is on the order of a molecular diameter, and $n$ depends on the nature of the interactions attracting the melt liquid to the solid, for example $n=3$ ($n=4$) corresponds to non-retarded (retarded) van der Waals interactions.  Minimizing $\Omega$ at fixed temperature, close to the bulk melting temperature $T_m$, total volume and chemical potential yields the film thickness, 

\begin{equation}
\label{eq:film_thickness}
d = \left[-\frac{(n-1)d_0^{n-1}\Delta \sigma}{\rho_l q_m}\right]^{1/n}t_r^{-1/n} \equiv \lambda_n t_r^{-1/n}, 
\end{equation}
where $t_r = (T_m - T)/T_m$ is the reduced temperature, $\rho_l$ is the liquid density, and $q_m$ is the latent heat of fusion \cite{Wettlaufer1995, Wettlaufer1996}. Moreover, the interfacial interactions create a pressure difference between the melted layer and the bulk solid \cite{Wettlaufer1995, Wettlaufer1996}, 

\begin{equation}
\label{eq:pressure_diff}
P_{l} - P_{s} = \Delta \sigma (n-1) {d_0}^{n-1} d^{-n} , 
\end{equation}
akin to the disjoining pressure of wetting dynamics \cite[e.g.,][]{deGennes}. Combining Eqs. \eqref{eq:film_thickness} and \eqref{eq:pressure_diff} a universal thermodynamic expression--independent of the nature of the interactions--lays bare the essential idea of premelting dynamics viz., 

\begin{equation}
\label{eq:pressure_temp}
P_l = P_s - \rho_l q_m t_r. 
\end{equation}
We fix $P_s$ by making contact with a reservoir and impose a temperature gradient parallel to the premelted surface. Therefore, as seen from Eq. \eqref{eq:pressure_temp}, $P_l$ increases with temperature and drives a flow in the film towards lower temperatures. Now, Eq. \eqref{eq:film_thickness} shows that the film thickness is a unique function of temperature and hence conservation of mass demands that a gradient in the volume flux must convert any liquid in excess of that value into solid as the liquid flows towards lower temperatures as shown in Fig. \ref{fig:schematic}. 

The basic dynamics of premelted films has been studied quantitatively in the case of ice, at both single crystal interfaces and in porous media, in part because of its geophysical implications \cite{Wettlaufer1996, Wilen1995, Rempel2004, Dash2006, Wettlaufer2006}, particularly in frost heave.  It has also been studied in argon \cite{Zhu2000}, benzene \cite{Taber1929} and helium \cite{Mizusaki1995}, which contract upon solidification.  

\section{Confinement effects \label{sec:powerlaw_viscosity}}

Liquids confined between solids of other materials behave differently than their bulk counterparts.  For example as a confined liquid approaches a thickness of a handful of molecular layers, the bulk melting and freezing transitions vanish and molecular mobilities take values intermediate between those found in bulk solid or bulk liquid \cite{SchickLesHouches, Evans1990, Hugo2001}.  Molecular hydrodynamics show a layering normal to walls that depends on the nature of the wall-fluid interactions, but the influence on rheology is complex, ranging from shear thinning to shear thickening, so that new dynamical behavior emerges when a liquid is confined in a narrow gap \cite{Koplik1995}. 

The dynamical consequences of the basic issue of how a premelted interface transitions from a disordered solid to a bulk liquid as the temperature increases have thus far not been examined quantitatively. Eq. \eqref{eq:film_thickness} shows that as the temperature becomes arbitrarily small and the film becomes arbitrarily thin. Clearly, the concept of a bulk liquid will break down at thicknesses greater than the value of $d_0$, with the detailed nature of the breakdown depending on the material under consideration. As the film thins, the ordering imposed on the liquid by the solid may eventually span the interfacially melted gap, imprinting properties that are intermediate between a liquid and a solid. For example, when transecting the bulk ice-water interface from the solid side, the density and diffusivity transition from the bulk solid to the bulk liquid values over tens of angstroms, depending on crystallographic orientation of the solid and the degree of disequilibrium \cite[e.g.,][and refs. therein]{Nada:1997, Nada:2000, Nada:2016}. Hence, we expect that as the temperature of a premelted film decreases, such structural changes may have dynamical consequences.  For example, \citet{Pittenger2001} have shown that the viscosity of quasiliquid water on ice can be several orders of magnitude larger than the bulk value, which may also influence the kinetic friction of ice surfaces \cite[e.g.,][]{Butt1998, Dash2003}.  However, whereas premelting dynamics have been studied in some detail for ice surfaces, confinement effects on dynamical properties have been more thoroughly studied in other materials.  

The flow of confined liquids cannot be understood by simple extrapolation of the bulk properties. A number of experimental studies have focused on the viscosity of nanometer scale liquid films. For example, in experiments wherein thin film water of nanometer scale is confined between different surfaces, \citet{Dhinojwala1997} and \citet{Major2006} have shown that the effective viscosity of the confined water ($\eta_{\rm eff} \approx 3 \times 10^{4}$ Pa$\cdot$s) is approximately $10^7$ times larger than the bulk viscosity of water ($\eta_{\rm bulk} = 8.6 \times 10^{-4}$ Pa$\cdot$s). 
\citet{Bureau2010} has shown that the effective viscosity of octamethylcyclotetrasiloxane (OMCTS) confined in molecularly thin films can increase from the bulk value by approximately $10^2$.  Moreover, the effective viscosity increased with the number of molecular layers as a power law. On the other hand \citet{Raviv2002} conclude from their shear experiments that surface bound aqueous salt solutions retain bulk characteristics even when compressed to nanometer scales.  

\citet{Granick1991} showed experimentally that the effective viscosity of a simple confined molecular liquid (dodecane) appears to diverge as the thickness decreases under 40 nm.  \citet{Zhu2003} measured a confinement enhancement of the viscosity of OMCTS between mica sheets by two orders of magnitude as the sheet distance decreased by 14 nm. \citet{Zhu2004} measured a confinement enhancement of the viscosity of alkane fluids between mica sheets by two orders of magnitude (factors of 5-6) as sheet distance decreased by 6 nm under quasi-steady (rapid) quenching.  Besides the rate, an important factor in such experiments is the degree of perfection of the mica surfaces.  

In summary, whilst the experimental setting spans a range of materials and confining surfaces, they provide a conceptual framework to treat the dynamical consequences of the enhanced solid-like structure as premelted films thin in terms of an enhancement of the viscosity of the layer.  Such an enhancement will suppress the volume flux under thermomolecular pressure gradients, thereby changing the solidification process shown schematically in Fig. \ref{fig:schematic}, at least as the flow extends into low temperatures. 

\begin{figure}[htbp]
\centering 
\includegraphics[scale=0.5]{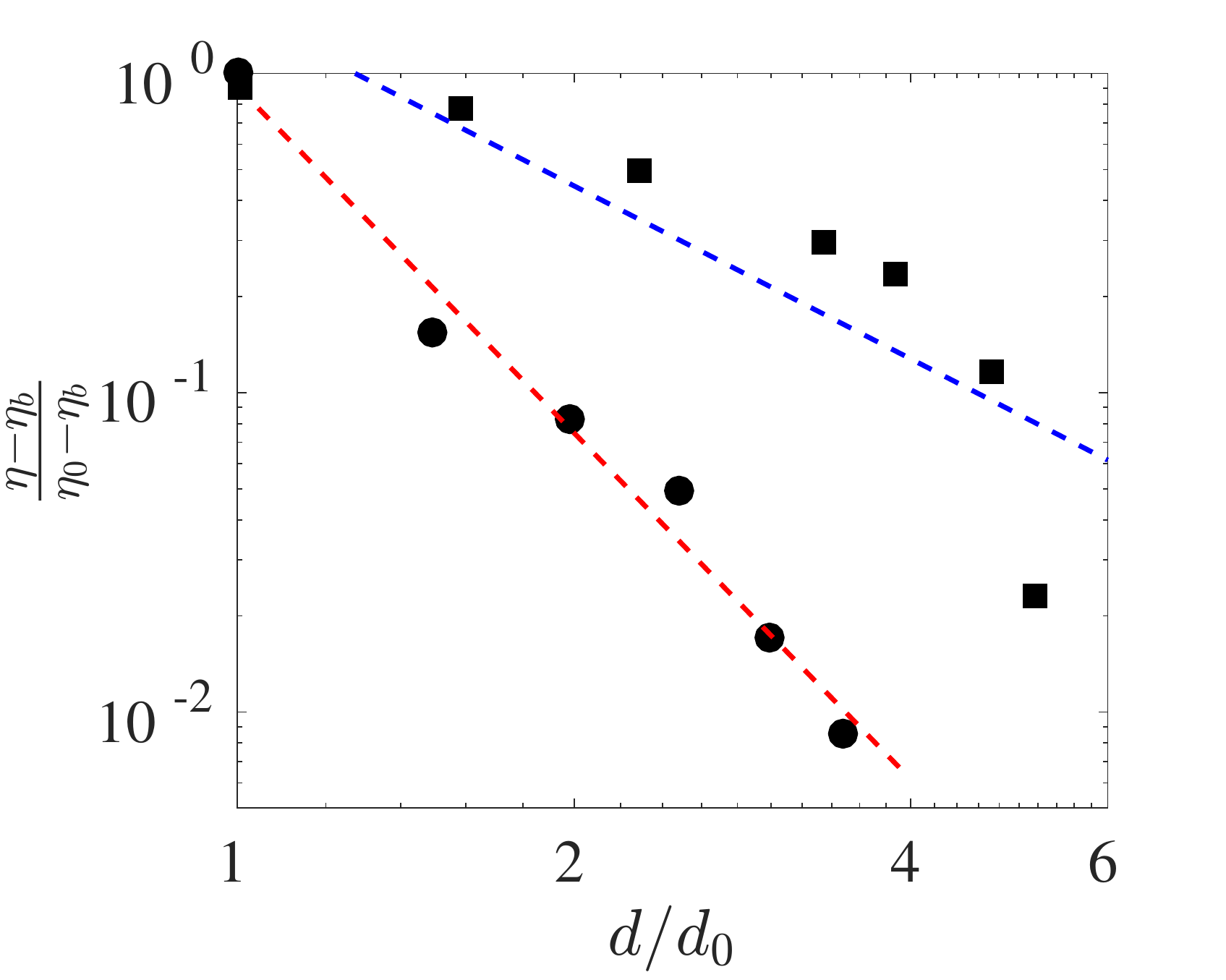} 
\caption{Least squares fit (dashed lines) of $\left(\eta-\eta_b\right)/\left(\eta_0-\eta_b\right) = \beta \left(d_0/d\right)^{\gamma}$ to the experimentally measured viscosity (filled symbols) from \cite{Becker2003} (circles; OMCTS), giving $\beta = 0.91, ~ \gamma = 3.60$ (red line), and \cite{Zhu2004} (squares; squalane), giving $\beta = 1.55, ~ \gamma = 1.80$ (blue line).} 
\label{fig:viscosity_fit}
\end{figure}

We understand the dynamics of premelted ice surfaces from theoretical and experimental studies of solidification in elastic capillary tubes \cite{Wettlaufer1995} and disks \cite{Wettlaufer1996}.  Here, we revisit the capillary tube problem using a proposed general power law relation for the dynamic viscosity of the melt liquid with the film thickness as 

\begin{equation}
\label{eq:powerlaw_viscosity_d}
\eta(d) = \eta_0 - \left(\eta_0 - \eta_b\right)\left[1 - \left(\frac{d_0}{d}\right)^{\gamma} \right], 
\end{equation}
where $\gamma > 0$ such that $\eta(d) \to \eta_b$ (bulk viscosity) as $d \to \infty$, and $\eta(d) \to \eta_0$ when $d \to d_0^{+}$, and the continuum approximation breaks down.  In order to generate a sense of the range of $\gamma$ for a given value of $\eta_r = \eta_0/\eta_b$, we perform a linear least squares fit of this model two sets of experimental data (Fig. \ref{fig:viscosity_fit}). This is preferred to fitting the confinement enhanced viscosity data from many individual experiments, such as those described above, for which we may have no information regarding the phenomenon of interfacial premelting.  We believe this avoids the risks associated with the introduction of more specific detail than can be justified by firm evidence.   

\section{Solidification in a capillary tube \label{sec:capillary_tube}}

Within the framework of a mathematical model we have previously studied the dynamics of premelted films in a capillary tube with elastic walls \cite{Wettlaufer1995}. Here we modify this theory by incorporation of the confinement effect through Eq. \eqref{eq:powerlaw_viscosity_d}, but in order to make our treatment reasonably self-contained we outline the key steps in the development of the key dynamical equation presently.  

As shown in Fig. 1 and described when discussing Eq. \eqref{eq:pressure_temp}, a temperature gradient is imposed parallel to the premelted surface will create a flow driven by a thermomolecular pressure gradient, 

\begin{equation}
\label{eq:thermo_pressure_gradient}
\nabla P_l = -\rho_l q_m \nabla t_r + \frac{\rho_l}{\rho_s}\nabla P_s, 
\end{equation}
where $P_s$ is the membrane-exerted external pressure on the solid.
Mass conservation in the thin film of thickness $d = r_2 - r_1$, between the solid of radius $r_1$ and the capillary tube of radius $r_2$, is 

\begin{equation}
\label{eq:mass_conservation}
\partial_t(\pi r_2^2) + \partial_x Q = 0, 
\end{equation}
where $Q$ is the volume flux, which for an annular lubrication flow through the film is 

\begin{equation}
\label{eq:volume_flux}
Q = -\frac{\pi}{6} \frac{r_2 d^3}{\eta}\left(\frac{\rho_l}{\rho_s}\partial_x P_s - \rho_l q_m \partial_x t_r\right). 
\end{equation}
Here, $\eta=\eta(d)$ is the dynamic viscosity of the melt given by Eq. \eqref{eq:powerlaw_viscosity_d}, and, as previously \cite{Wettlaufer1995}, we assume that the capillary wall exerts a linear elastic hoop stress upon the solid given by 

\begin{equation}
\label{eq:linear_elasticity}
P_s = k(r_2 - r_0), 
\end{equation}
where $k$ is a constant and $r_0$ is the undeformed radius, viz., $r_2(x,t = 0) = r_0(x)$. 

For constant bulk viscosity in the case of both transient and steady state thermal fields, similarity solutions of the capillary radius were obtained \cite{Wettlaufer1995}, but here we are interested in the latter case and thus take a constant temperature gradient $\mathcal{G}$, such that the reduced temperature is $t_r = \mathcal{G}x/T_m$. Thus, in this temperature field, using Eq. \eqref{eq:film_thickness} in Eq. \eqref{eq:powerlaw_viscosity_d}, the dynamic viscosity becomes a function of position viz., 

\begin{equation}
\label{eq:powerlaw_viscosity_x}
\eta(x) = \eta_0 - \left(\eta_0 - \eta_b\right) \left[1 - \left\{\frac{d_0}{\lambda_n}\left(\frac{\mathcal{G}}{T_m}\right)^{1/n}\right\}^{\gamma} x^{\gamma/n}\right], 
\end{equation}
and combined with Eqs. \eqref{eq:mass_conservation} - \eqref{eq:linear_elasticity}, mass conservation can be written solely in terms of the capillary radius as 

\begin{eqnarray}
& & \partial_t r_2 -  \frac{\rho_l k \lambda_n^3}{12 \rho_s}\left(\frac{T_m}{\mathcal{G}}\right)^{3/n}\partial_x \Bigg[\frac{x^{-3/n}}{\eta(x)} \partial_x \bigg(r_2 - r_0 \nonumber \\ 
\label{eq:mass_conservation_1}
& & ~~~~~~~~~~~~~~~~~~~~~~~~~~~~~~~~~~ - \frac{\rho_s q_m \mathcal{G}}{k T_m} x \bigg)\Bigg] = 0. 
\end{eqnarray}
Whence, the capillary deformation can be obtained by solving Eq. \eqref{eq:mass_conservation_1} subject to the following boundary and initial conditions, 

\begin{subequations}
\begin{align}
& r_2 - r_0 = 0 ~ (x = 0) ~ \text{and} ~ r_2 - r_0 = 0 ~ (x \to \infty), \label{eq:BCs} \\
& r_2(x,t=0) = r_0(x), \label{eq:IC} \\ & \text{and} \nonumber \\
& \partial_x (r_2 - r_0) = \frac{\rho_s q_m \mathcal{G}}{k T_m} ~ (x = 0), \label{eq:NC}
\end{align}
\end{subequations}
Eqs. \eqref{eq:BCs} express no deformation at the bulk solid-liquid interface and as $d \rightarrow d_0^+$, while Eq. \eqref{eq:NC} is a necessary condition to remove the singularity at $x = 0$. 

\subsection{Non-dimensionalization \label{subsec:nondim}}
We render the above system of equation dimensionless with the following scaling 

\begin{subequations}
\begin{align}
& \tilde{x} = \frac{x}{X_0}, ~~~ \tilde{t} = \frac{t}{X_0^{2+3/n}/\Gamma_n}, ~~~ \tilde{r} = \frac{r_2 - r_0}{\alpha X_0}, \label{eq:dimensionless_variables_powerlaw_force}  \\
& \tilde{\eta}(\tilde{x}) =  \frac{ \eta_0 - \eta }{ \eta_0 - \eta_b } = \left[ 1 - \left\{ \frac{d_0} {\lambda_n} \left( \frac{ \mathcal{G} X_0 }{ T_m } \right)^{1/n} \right\}^{\gamma} \tilde{x}^{\gamma/n} \right] \nonumber \\
& ~~~~~~~~~~~~~~~~~~~~ = \left[1 - d_{\gamma, n} \tilde{x}^{\gamma/n}\right], \label{eq:dimensionless_viscosity} 
\end{align}
\end{subequations}
wherein
\begin{equation}
\Gamma_n = \frac{\rho_l k \lambda_n^3}{12 \eta_b \rho_s}\left(\frac{T_m}{\mathcal{G}}\right)^{3/n}, ~~~ \alpha = \frac{\rho_s q_m \mathcal{G}}{k T_m}, 
\end{equation}
from which we obtain 

\begin{subequations}
\begin{align}
& \partial_{\tilde{t}} \tilde{r} + \partial_{\tilde{x}}\left[-\frac{\tilde{x}^{-3/n}}{\eta_1(\tilde{x})} \left(\partial_{\tilde{x}} \tilde{r} - 1\right) \right] = 0, \label{eq:mass_conservation_powerlaw_force} \\
& \eta_1(\tilde{x}) = \eta_r - \left(\eta_r - 1\right) \tilde{\eta}(\tilde{x}) = 1 + \left(\eta_r - 1\right)d_{\gamma,n}\tilde{x}^{\gamma/n}, \label{eq:powerlaw_visosity1} \\
& \tilde{r} = 0 ~ (\tilde{x} = 0, 1), \label{eq:nondimBCs} \\
& \tilde{r} = 0 ~ (\tilde{t} = 0), \label{eq:nondimIC} \\ & \text{and} \nonumber \\
& \partial_{\tilde{x}} \tilde{r} - 1 = 0, \label{eq:nondimNC}
\end{align}
\end{subequations}
Here, $X_0$ is a length scale chosen based on the temperature gradient. For clarity of notation, the dimensionless variables are now denoted without the tilde overscripts, and the dimensional variables are explicitly mentioned with their respective SI units. 

Expanding the flux gradient, $\displaystyle \partial_x \mathcal{Q} \equiv \partial_x \left[-\frac{x^{-3/n}}{\eta_1(x)} \left(\partial_x r - 1\right)\right]$, we rewrite Eq. \eqref{eq:mass_conservation_powerlaw_force} in the form of an advection-diffusion equation with a source, $ \mathcal{F}_n(x)$, as 

\begin{equation}
\label{eq:general_ADE_sourc_powerlaw_force}
\partial_t r = \mathcal{D}_n(x)\partial_{xx} r + \mathcal{U}_n(x)\partial_x r + \mathcal{F}_n(x), 
\end{equation}
in which
\begin{subequations}
\begin{align}
& \mathcal{D}_n(x) = \frac{x^{-3/n}}{\eta_1(x)} > 0, \label{eq:D_powerlaw_forces} \\
& \mathcal{U}_n(x) = - \frac{x^{-(1+3/n)}}{\eta_1^2(x)}\left[\left(\frac{3}{n} + \frac{\gamma}{n}\right)\eta_1(x) - \frac{\gamma}{n}\right] < 0, \label{eq:U_powerlaw_forces}\\&\text{and} \nonumber \\
& \mathcal{F}_n(x) = \frac{x^{-(1+3/n)}}{\eta_1^2(x)}\left[\left(\frac{3}{n} + \frac{\gamma}{n}\right)\eta_1(x) - \frac{\gamma}{n}\right] > 0. \label{eq:F_powerlaw_forces}
\end{align}
\end{subequations}

A key difference between the previous treatment \cite{Wettlaufer1995} and the one here is that the proximity effect,  treated through the viscosity model of  Eq. \eqref{eq:powerlaw_visosity1},  introduces an inhomogeneity so that Eq. \eqref{eq:general_ADE_sourc_powerlaw_force} does not have a general similarity solution.  However, we note that for $\eta_1(x) = 1$ (i.e., $\eta = \eta_b$, from Eqs. \eqref{eq:dimensionless_viscosity} and \eqref{eq:powerlaw_visosity1}), Eq. \eqref{eq:mass_conservation_1} simplifies to 

\begin{equation}
\label{eq:bulk_viscosity_alln}
\partial_{t} r_2 + \Gamma_n \partial_{x} \left[ x^{-3/n} \left\{ \alpha - \partial_{x} (r_2 - r_0) \right\} \right] = 0, 
\end{equation}
which is satisfied by the similarity solution, 

\begin{equation}
\label{eq:old_similarity_sol} 
r_2 - r_0 = \alpha \left(\Gamma_n t\right)^{n/(2n+3)}g\left(\zeta\right), 
\end{equation}
with similarity variable 

\begin{equation}
\label{eq:old_similarity_variable} 
\zeta = x \left(\Gamma_n t\right)^{-n/(2n+3)}. 
\end{equation} 
One obtains the similarity solution $g(\zeta)$ by solving a boundary value problem of a family of second order, ordinary differential equations, 

\begin{subequations}
\begin{align}
& g - \zeta g^{\prime} = \left(2 + \frac{3}{n}\right) \left[\zeta^{-3/n}\left(g^{\prime} - 1\right)\right]^{\prime}, \label{eq:ODE_g_powerlaw_force} \\
& g = 0 ~~~~ (\zeta  = 0) ~~ \text{and} ~~ g \to 0 ~~~~ (\zeta \to \infty), \label{eq:BCs_similarity_powerlaw_force} 
\end{align}
\end{subequations}
where the primes denote ${\rm d}/{\rm d}\zeta$, supplemented by the necessary condition 

\begin{equation}
\label{eq:essential_condition}
g^{\prime} = 1 ~~~~ (\zeta = 0)
\end{equation}
that there is no singularity at the bulk solid-liquid interface. 

\section{Results \label{sec:results}} 

\subsection{Analytical results \label{subsec:AR}} 

\subsubsection{Approximation of $\gamma_{\rm crit}(\eta_r, d_0, n)$ and $\gamma_{\rm sat}(\eta_r, d_0, n)$ \label{subsubsec:analytical}} 

Clearly, the nature of the solution to Eq. \eqref{eq:general_ADE_sourc_powerlaw_force} depends on the proximity effect embodied in the behavior of $\eta_1(x)$. For example, we obtain the similarity solution \eqref{eq:old_similarity_sol} either for $\eta_r = 1$ or when $\eta_1(x) = 1 + \left(\eta_r - 1\right)d_{\gamma,n}x^{\gamma/n} \sim \mathcal{O}(1)$, for all $x$, and $\gamma \geq \gamma_{\rm sat}(\eta_r, d_0, n)$.
Here, the subscript `sat' denotes saturation of the viscosity, $\eta(d),$ to the bulk viscosity, $\eta_b$. 

Now, we estimate $\gamma_{\rm sat}(\eta_r, d_0, n)$ and $\gamma_{\rm crit}(\eta_r, d_0, n)$ (described below) from  $\eta_1(x)$. For $0 \leq x \leq 1$, when $\eta_r d_{\gamma,n} \sim \mathcal{O}(1)$, then 

\begin{equation}
1 + \left(\eta_r - 1\right)d_{\gamma,n}x^{\gamma/n} \sim \mathcal{O}(1), 
\end{equation}
which yields, 
\begin{equation}
\label{eq:gamma_sat}
\gamma_{\rm sat}(\eta_r, d_0, n) = -\frac{\ln \eta_r}{\ln\left(d_0/\lambda_n\right) + \ln(\mathcal{G}X_0/T_m)/n}. 
\end{equation}

On the other hand, when $\eta_1(x) = 1 + \left(\eta_r - 1\right)d_{\gamma,n}x^{\gamma/n} \sim \mathcal{O}(x^{\gamma/n}), ~ \forall x$, the capillary deformation is given by 

\begin{equation}
\label{eq:new_ODE_g_powerlaw_force}
g_{\gamma} - \zeta_{\gamma} g_{\gamma}^{\prime} = \left(2 + \frac{3-\gamma}{n}\right) \left[\zeta_{\gamma}^{-\frac{3-\gamma}{n}}\left(g_{\gamma}^{\prime} - 1\right)\right]^{\prime}.  
\end{equation}
Subject to conditions, Eqs. \eqref{eq:BCs_similarity_powerlaw_force} and \eqref{eq:essential_condition}, 

\begin{equation}
\label{eq:new_similarity_sol}
g_{\gamma}(\zeta_{\gamma}) = (r_2 - r_0) \alpha^{-1} \left[ \frac{\Gamma_n t}{(\eta_r-1)d_{\gamma, n}} \right]^{-n/(2n+3+\gamma)}, 
\end{equation}
with similarity variable 

\begin{equation}
\label{eq:new_similarity_variable}
\zeta_{\gamma} = x\left[\frac{\Gamma_n t}{(\eta_r-1)d_{\gamma, n}}\right]^{-n/(2n+3+\gamma)}. 
\end{equation}

When $\eta_1(x) = 1 + \left(\eta_r - 1\right)d_{\gamma,n}x^{\gamma/n}$ exhibits both $\mathcal{O}(1)$ and $\mathcal{O}(x^{\gamma/n})$ behaviors locally in $0 \leq x \leq 1$, there is a length scale imposed upon the system, which preempts the existence of a similarity solution.   One can show that 

\begin{equation}
1 + \left(\eta_r - 1\right)d_{\gamma,n}x^{\gamma/n} \sim \mathcal{O}(1), ~ \text{at} ~ x = 0^+, 
\end{equation}
and 

\begin{equation}
1 + \left(\eta_r - 1\right)d_{\gamma,n}x^{\gamma/n} \sim \mathcal{O}(x^{\gamma/n}), ~ \text{for} ~ x > 0^+, 
\end{equation}
when $\eta_r d_{\gamma,n}\epsilon^{\gamma/n} \sim \mathcal{O}(1)$ for $0 < \epsilon \ll 1$, which yields, 

\begin{equation}
\label{eq:gamma_crit}
\gamma_{\rm crit}(\eta_r, d_0, n) = -\frac{\ln \eta_r}{\ln\left(d_0/\lambda_n\right) + \left[\ln(\mathcal{G}X_0/T_m) + \ln \epsilon\right]/n}. 
\end{equation}

When the solution of Eq. \eqref{eq:mass_conservation_powerlaw_force} for the proximity viscosity model with $\gamma \in (\gamma_{\rm crit}, \gamma_{\rm sat})$ intersects the solution corresponding to the bulk viscosity model at $x=x_{\rm cross}$, then, upon substitution of $\epsilon = x_{\rm cross}$ into Eq. \eqref{eq:gamma_crit}, one obtains the corresponding value of $\gamma$. Since we have no {\em a priori} information about $x_{\rm cross}$, we can not provide an analytic expression for it. However, we can use numerical solutions to compute $x_{\rm cross}$ {\em a posteriori}. Thus, one obtains $\gamma_{\rm sat}$ from $\gamma_{\rm crit}$ in the limit $\epsilon \to 1$. 

Taking the limit $\gamma \to 0^+$, $\eta \to \eta_0$ in Eq. \eqref{eq:dimensionless_viscosity}, the viscosity of the liquid film is independent of the film thickness. This is analogous to the previous model  \cite{Wettlaufer1995}, but with the viscosity increased by a factor $\eta_r$, and we obtain the similarity solution 

\begin{equation}
\label{eq:new_similarity_sol_gamma0}
g_{\gamma \to 0^+} = r \alpha^{-1}\left(\frac{\Gamma_n t}{\eta_r}\right)^{-n/(2n+3)} = g(\zeta) \eta_r^{n/(2n+3)}, 
\end{equation}
with similarity variable 

\begin{equation}
\label{eq:new_similarity_variable_gamma0}
\zeta_{\gamma \to 0^+} = x\left(\frac{\Gamma_n t}{\eta_r}\right)^{-n/(2n+3)} = \zeta \eta_r^{n/(2n+3)}, 
\end{equation}
which satisfies \eqref{eq:ODE_g_powerlaw_force}-\eqref{eq:BCs_similarity_powerlaw_force}, and the necessary condition \eqref{eq:essential_condition}. 

\subsection{Numerical results \label{subsec:NR}} 

We discretized Eq. \eqref{eq:general_ADE_sourc_powerlaw_force} using the standard second-order finite difference formulae and integrate in time using the semi-implicit Crank-Nicolson method \cite{LeVeque2007a, Toppaladoddi2017}. We use uniform grid spacing $\Delta x$ and $\Delta t$ for the space and time, respectively (see the Supplemental Material for the numerical scheme \cite{SM}). 

\begin{figure*}[htbp]
\centering 
(a) \hspace{3in} (b) \\
\includegraphics[scale=0.5]{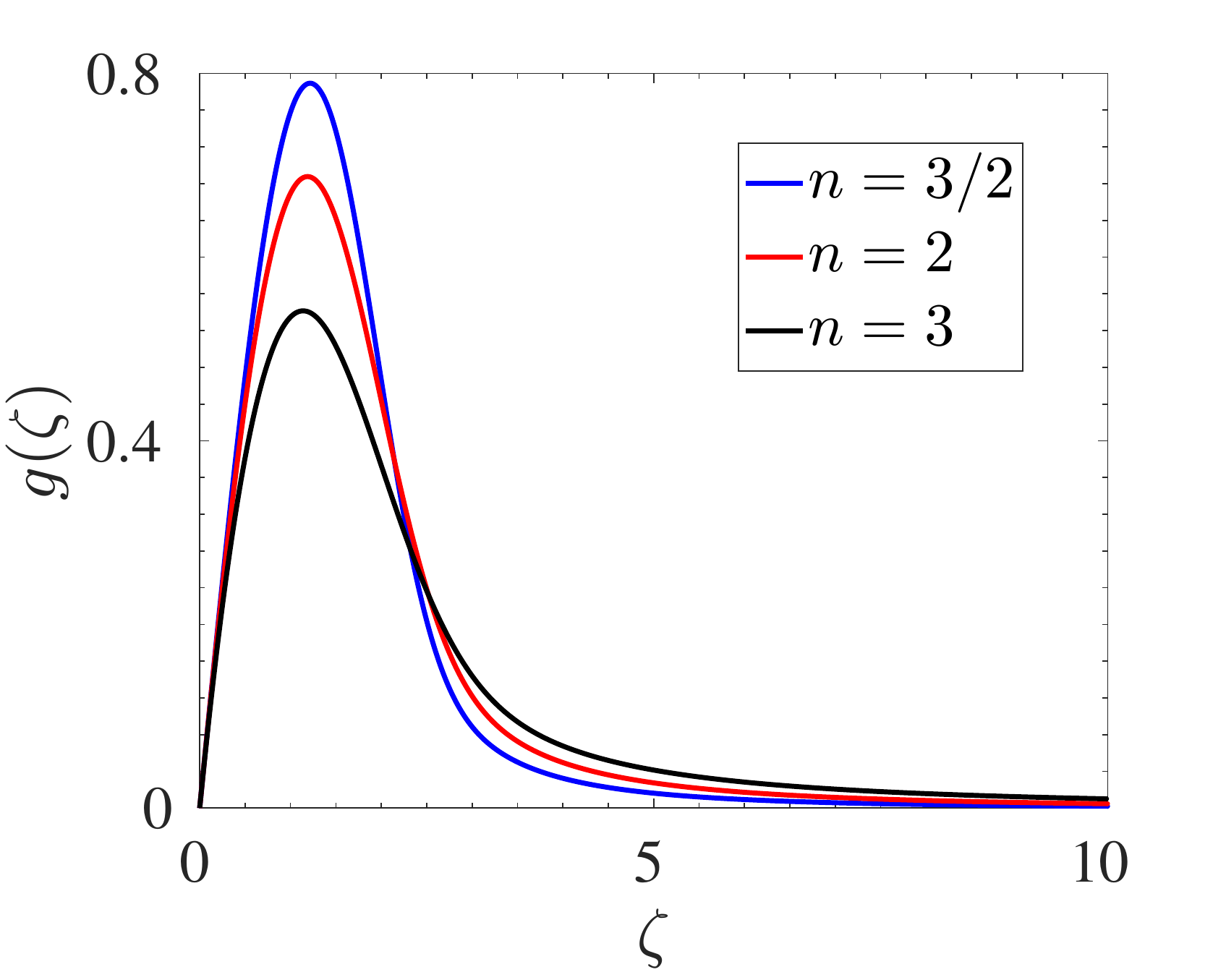} 
\includegraphics[scale=0.5]{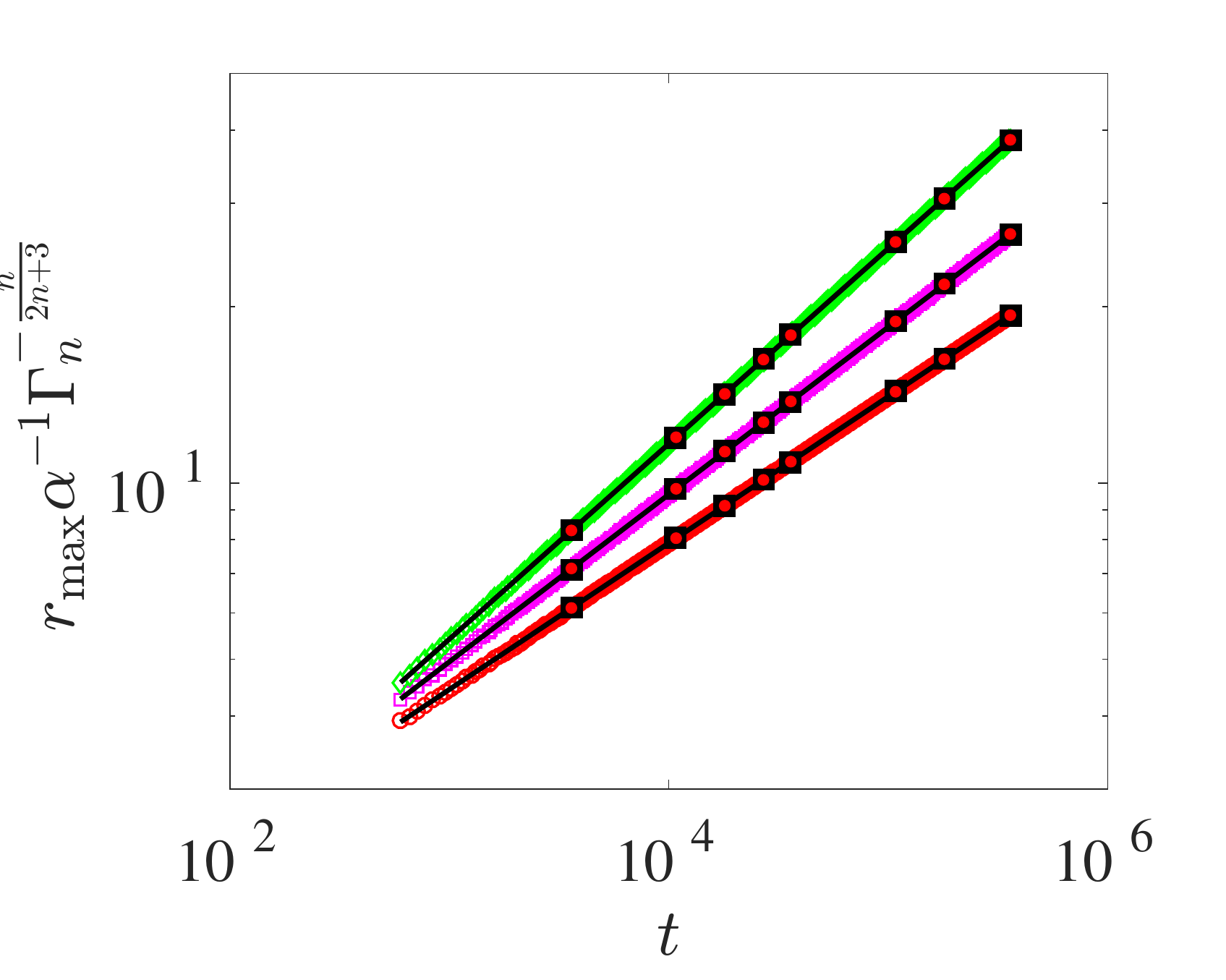} 
\caption{(a) Similarity solution, Eq. \eqref{eq:old_similarity_sol}, for the short and long ranged electrostatic ($n = 3/2$ and $n = 2$) and nonretarded van der Waals ($n = 3$) interactions. (b) For each type of interaction (i.e., each $n$), a linear least square fit gives a power law relation for the maximum capillary deformation, $r_{\rm max}(t) \propto t^{n/(2n+3)}$, as given by Eq. \eqref{eq:old_similarity_sol}. The $t^{1/4}$, $t^{2/7}$, and $t^{1/3}$ power laws are for $n = 3/2$ (red), $n = 2$ (magenta), and $n = 3$ (green) respectively. The solid lines correspond to $g_{{\rm max}, n}t^{n/(2n+3)}$ and the filled symbols correspond to the parameters for dodecane for $d_0 = 1$ \AA  ~(black squares) and $d_0 = 10$ \AA ~ (red dots).} 
\label{fig:similarity_sol}
\end{figure*}

The similarity solution, Eq. \eqref{eq:old_similarity_sol}, is constructed from the numerical solution of Eq. \eqref{eq:general_ADE_sourc_powerlaw_force} for the bulk viscosity model and is plotted in Fig. \ref{fig:similarity_sol}(a) for the short and long ranged electrostatic ($n = 3/2$ and $n = 2$) and nonretarded van der Waals ($n = 3$) interactions. The self-similar structure of the solution is obtained with $g_{\rm max, 3} \approx 0.5415$ at $\zeta_{\rm max, 3} \approx 1.1341$;  $g_{\rm max, 2} \approx 0.6877$ at $\zeta_{\rm max, 2} \approx 1.1853$ and $g_{\rm max, 3/2} \approx 0.7894$ at $\zeta_{\rm max, 3/2} \approx 1.2162$. From Eqs. \eqref{eq:dimensionless_variables_powerlaw_force}, \eqref{eq:old_similarity_sol} and \eqref{eq:old_similarity_variable}, it can be shown that $r_{\rm max} = \alpha g_{{\rm max}, n} \Gamma_n^{n/(2n + 3)} t^{n/(2n + 3)}$.  Furthermore, to demonstrate the robustness of the numerical method used, we calculated $r_{\rm max} \alpha^{-1} \Gamma_n^{-n/(2n + 3)}$ as a function of $t$ from the numerical solutions, which are plotted in Fig. \ref{fig:similarity_sol}(b). For any given $n$, the scaling behavior of the similarity solution is independent of the molecular diameter ($d_0$) and $\lambda_n$. In other words, the similarity solution is characterized by a single parameter, $n$. For the non-retarded van der Waals interactions ($n = 3$), we recover the results of \citet{Wettlaufer1995} (see their Eqs. (51)-(56)).

The results from the numerical computations presented below were obtained using a temperature gradient of $\mathcal{G} = 1$ K/m (or $10^2$ K/m), a final (dimensional) time of integration $T_f = 10^2$ h, and (a) $n = 3$, $\eta_r = 10^2, 10^3, 10^4, 10^5, 10^6, 10^7$, $d_0 = 8$ \AA, $\lambda_3 = 5.3947$ \AA; (b) $n = 2$, $\eta_r = 10^7$, $d_0 = 8$ \AA, $\lambda_2 = 5.3947$ \AA. The range of power law exponent, $\gamma$, explored is tabulated in Appendix \ref{app:parameters_water}.  

\begin{figure}[htbp]
\centering 
\includegraphics[scale=0.5]{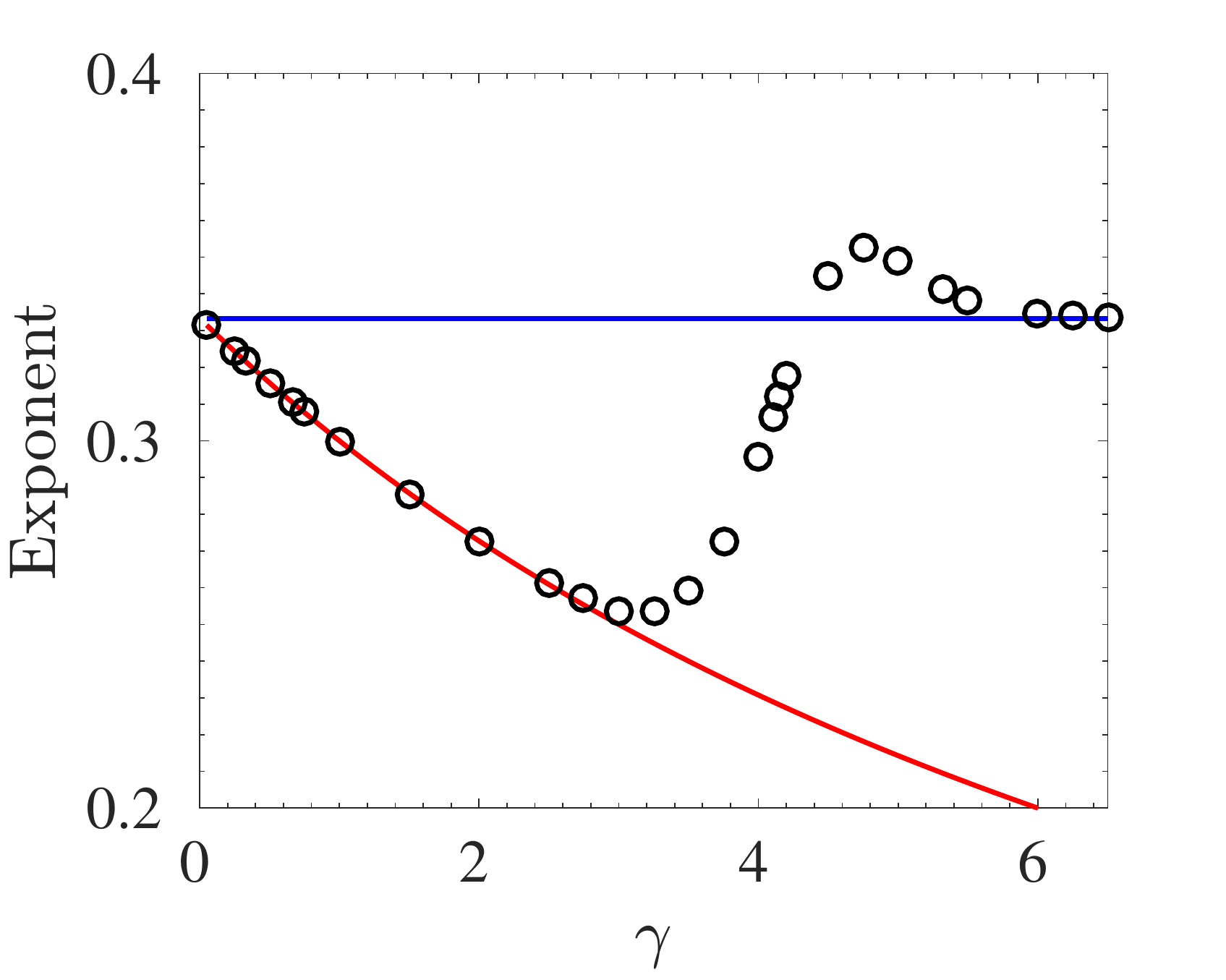} 
\caption{For $n = 3, ~ d_0 = 8$ \AA, $\mathcal{G} = 1$ K/m, $\eta_r = 10^7$ and each $\gamma$, a linear least squares fit is used to obtain a power law relation, $r_{\rm max} \propto t^{b(\gamma, \eta_r, d_0, n)}$. Circles represent $b(\gamma, \eta_r, d_0, n)$ from the linear least squares fit. The red and blue lines represent the exponents $n/(2n + 3 +\gamma)$ and $n/(2n + 3)$, respectively, corresponding to the similarity solutions \eqref{eq:old_similarity_sol} and \eqref{eq:new_similarity_sol}.} 
\label{fig:exponent}
\end{figure}

\begin{figure*}[hbtp]
\centering 
(a) \hspace{3in} (b) \\ 
\includegraphics[scale=0.5]{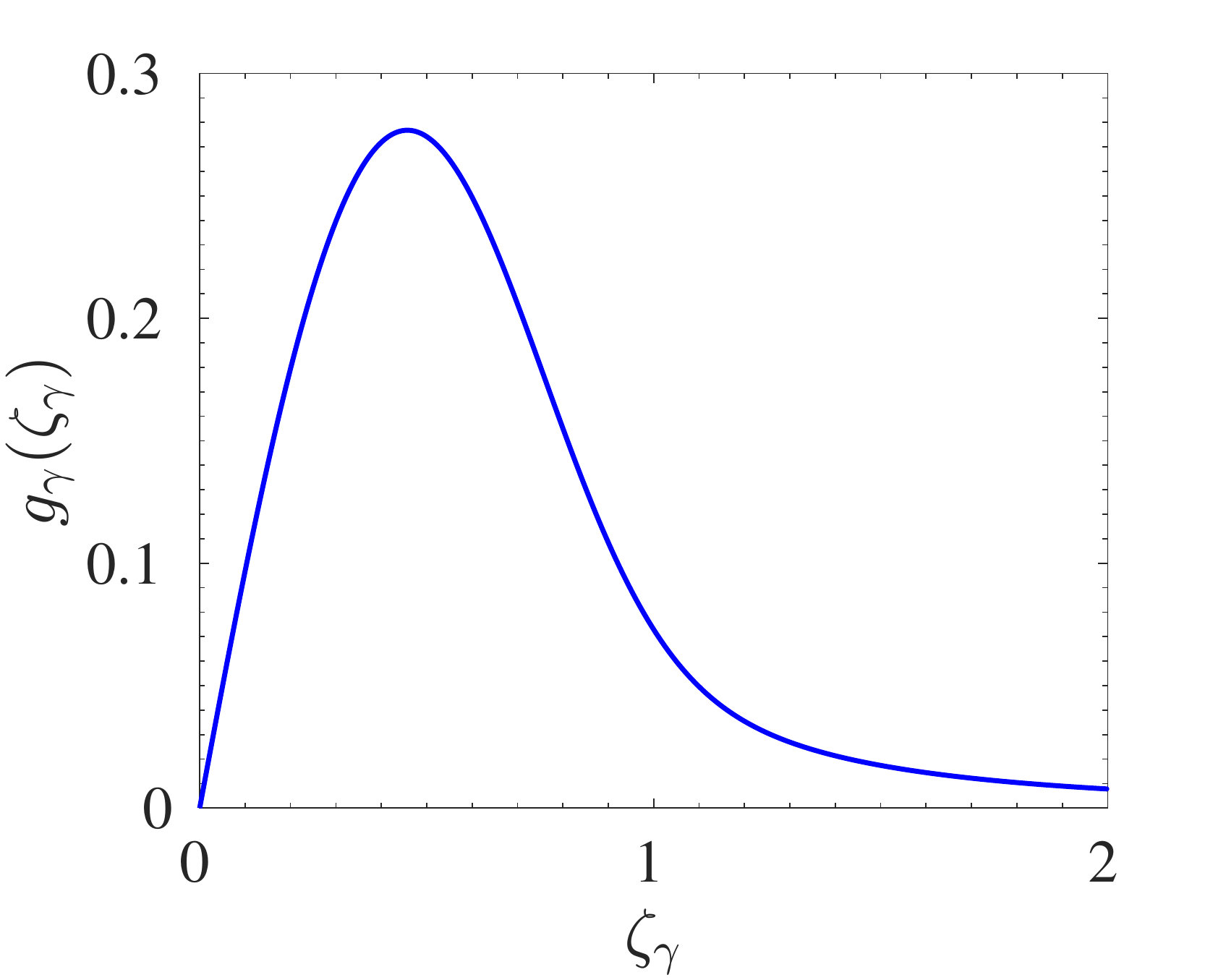} 
\includegraphics[scale=0.5]{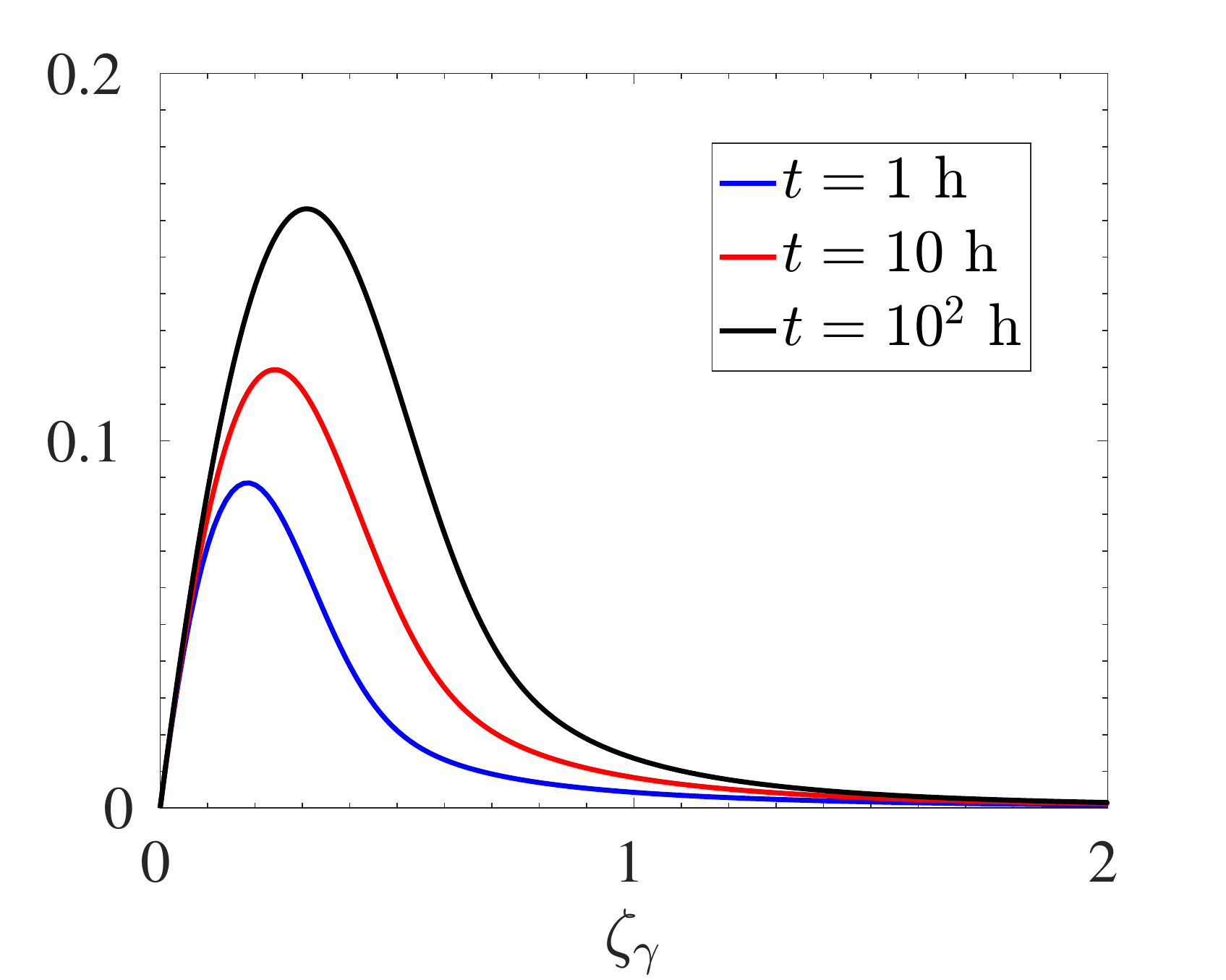} 
\caption{For $\eta_r = 10^7$, the deformation in presented in terms of the similarity variables, viz., Eq. \eqref{eq:new_similarity_variable}: (a) $g_2(\zeta_2)$ is self-similar, but (b) $g_{4.75}(\zeta_{4.75})$ is not self-similar 
(The legend in the inset corresponds to the time in dimensional units.). 
Note the difference in the vertical scale.} 
\label{fig:different_similarity_sol}
\end{figure*}

\subsubsection{$\gamma_{\rm crit}(\eta_r, d_0, n)$ and $\gamma_{\rm sat}(\eta_r, d_0, n)$} 

For given values of $\gamma, ~ \eta_r, ~ d_0$ and $n$ (hence $\lambda_n$), we compute the maximum deformation, $r_{\rm max}(t; \gamma, \eta_r, d_0, n)$, as a function of time, $t$. For each $\gamma$, a linear least square fit is used to obtain a power law: 
\begin{equation}
\label{eq:fit_powerlaw}
r_{\rm max}(t; \gamma, \eta_r, d_0, n) = a(\gamma, \eta_r, d_0, n)t^{b(\gamma, \eta_r, d_0, n)}. 
\end{equation}
Fig. \ref{fig:exponent} shows $b(\gamma, \eta_r, d_0, n)$ for $\eta_r = 10^7, ~ d_0 = 8$ \AA, and $n = 3$ along with the exponents $n/(2n + 3)$ and $n/(2n + 3 + \gamma)$ of the two similarity solutions \eqref{eq:old_similarity_sol} and \eqref{eq:new_similarity_sol}, respectively. We show that $b(\gamma, \eta_r, d_0, n)$ coincides with $n/(2n + 3 + \gamma)$ for $0 < \gamma \lesssim 2.75$, and that with $n/(2n + 3)$ for $\gamma \gtrsim 6.25$. This is in accordance with our analytical argument above concerning the existence of two different similarity solutions depending upon the power-law exponent $\gamma$. For a more quantitative comparison between the analytical and numerical results, we compute $\gamma_{\rm crit}(\eta_r, d_0, n)$ and $\gamma_{\rm sat}(\eta_r, d_0, n)$ numerically as 

\begin{widetext}
\begin{subequations}
\begin{align}
& \gamma_{\rm crit}(\eta_r, d_0, n) = \max \left\{ \gamma > 0 : \left| b(\gamma, \eta_r, d_0, n) - \frac{n}{2n+3+\gamma} \right| \leq 10^{-4} \right\}, \label{eq:gammacrit_numeric} \\ 
& \gamma_{\rm sat}(\eta_r, d_0, n) = \min \left\{ \gamma > 0 : \left| b(\gamma + \delta, \eta_r, d_0, n) - \frac{n}{2n+3}\right| \leq 10^{-4}, ~ \forall \delta \geq 0 \right\}. \label{eq:gammasat_numeric} 
\end{align}
\end{subequations}
\end{widetext}
The numerically and analytically determined $\gamma_{\rm crit}$ and $\gamma_{\rm sat}$ are shown in Table \ref{tab:critical_saturation_gamma} for $d_0 = 8$ \AA: (a) $n = 3$ and $\eta_r = 10^5, 10^6, 10^7$; (b) $n = 2$ and $\eta_r = 10^7$. 

\begin{table}
\centering
\begin{tabular}{|cc|cc|cc|} \hline
\multicolumn{2}{|c|}{~} & \multicolumn{2}{|c|}{Numerical} & \multicolumn{2}{|c|}{Analytical} \\ 
$n$ & $\eta_r$ & $\gamma_{\rm crit}$ & $\gamma_{\rm sat}$ & $\gamma_{\rm crit}$ & $\gamma_{\rm sat}$ \\ \hline
2 & $10^7$ & 2 & 4.75 & 2.09 & 4.12 \\ \hline 
3 & $10^5$ & 1.75 & 4.75 & 2.29 & 4.65 \\ \hline 
3 & $10^6$ & 2.25 & 5.5 & 2.75 & 5.58 \\ \hline 
3 & $10^7$ & 2.75 & 6.25 & 3.21 & 6.51 \\ \hline 
\end{tabular}
\caption{Numerical and analytical values of $\gamma_{\rm crit}$ and $\gamma_{\rm sat}$.} 
\label{tab:critical_saturation_gamma}
\end{table}

\begin{figure*}[hbtp]
\centering 
(a) \hspace{3in} (b) \\ 
\includegraphics[scale=0.5]{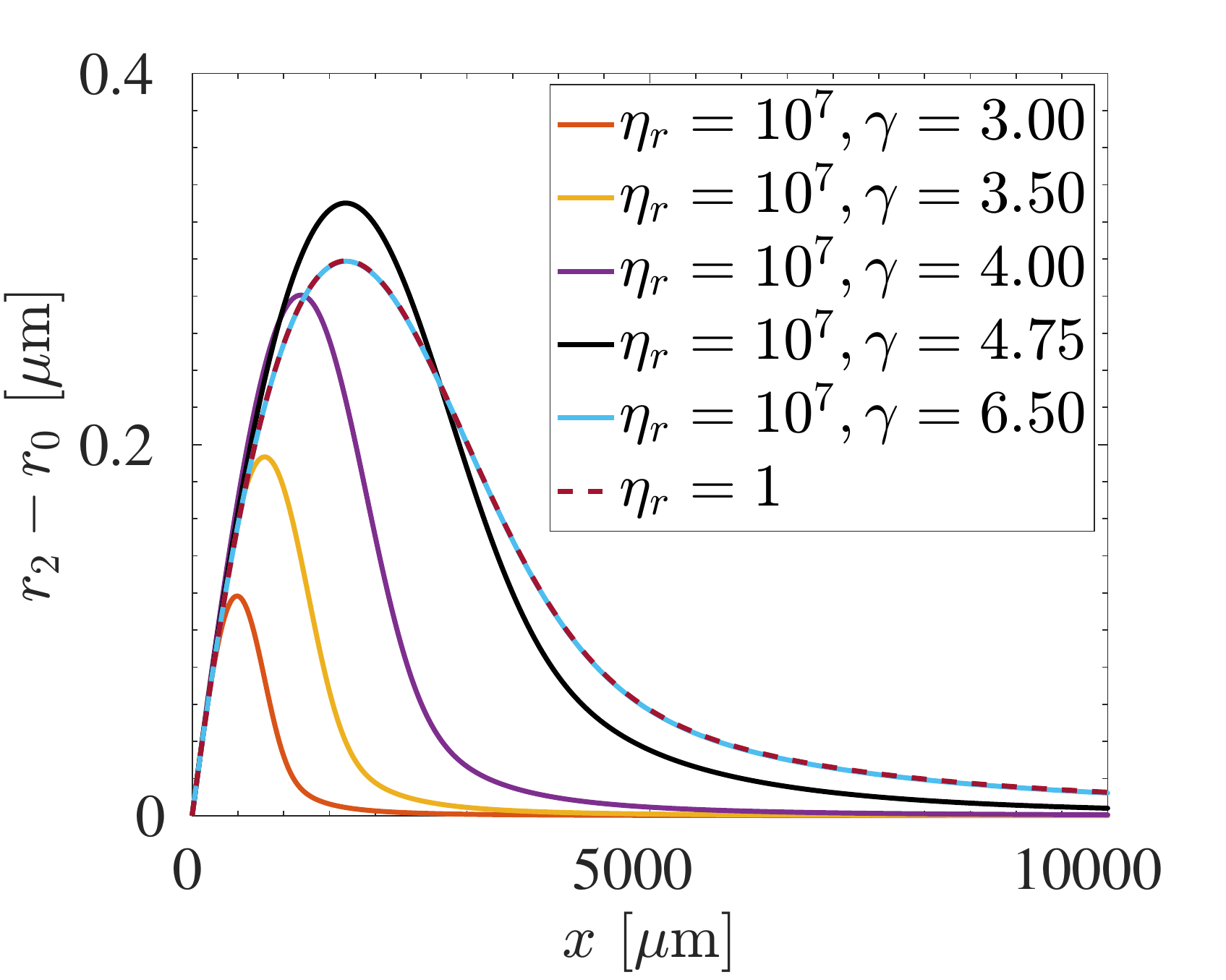} 
\includegraphics[scale=0.5]{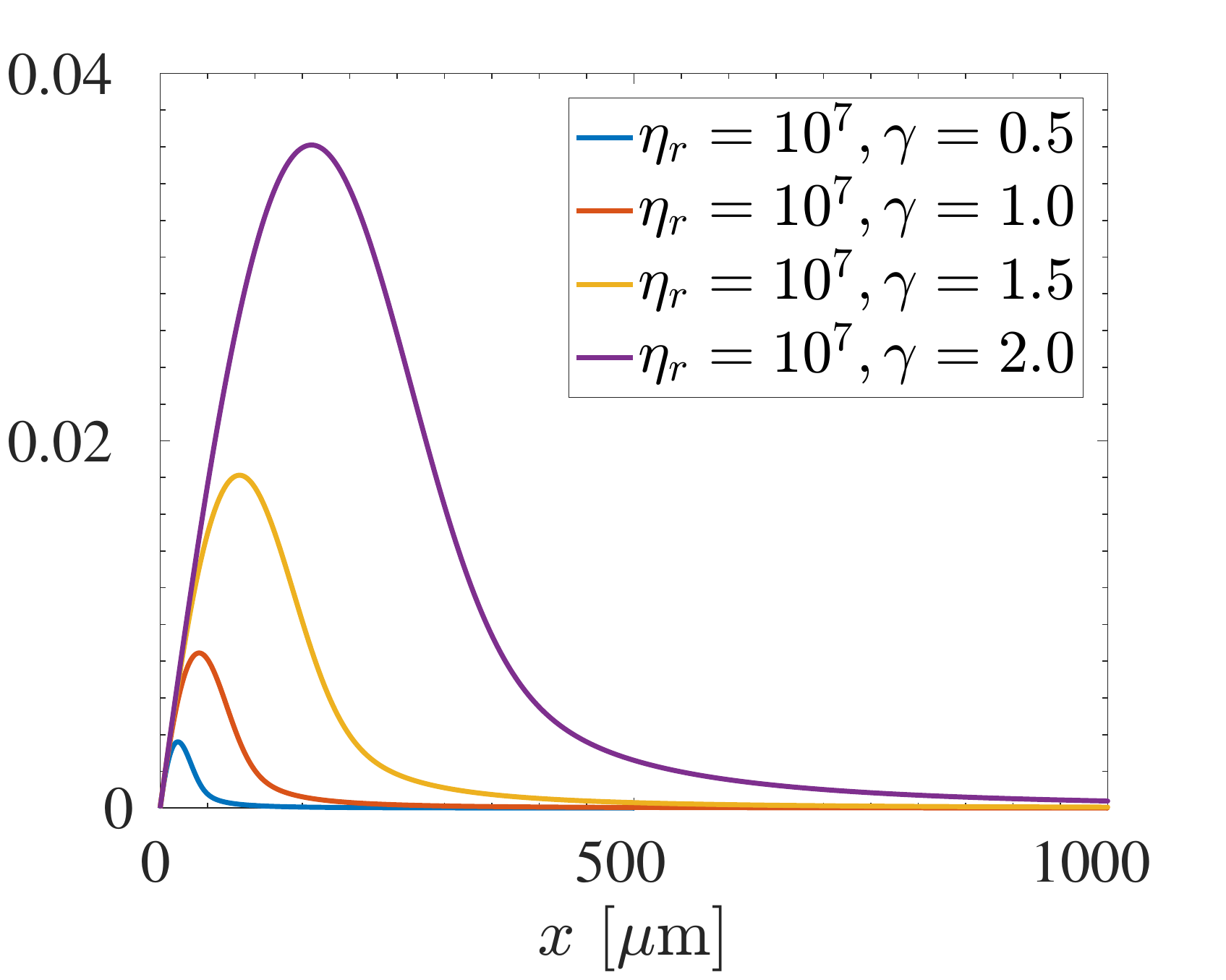} 
\caption{Dimensional deformation as a function of position at dimensional time $10^2$ h for the parameters shown in the insets, and $\mathcal{G} = 1$ K/m, $n = 3$, $d_0 = 8$ \AA. Note the difference in the vertical scales.} 
\label{fig:etar1e7_d0_8angstrom_n3}
\end{figure*}

\begin{figure}[hbtp]
\centering 
\includegraphics[scale=0.5]{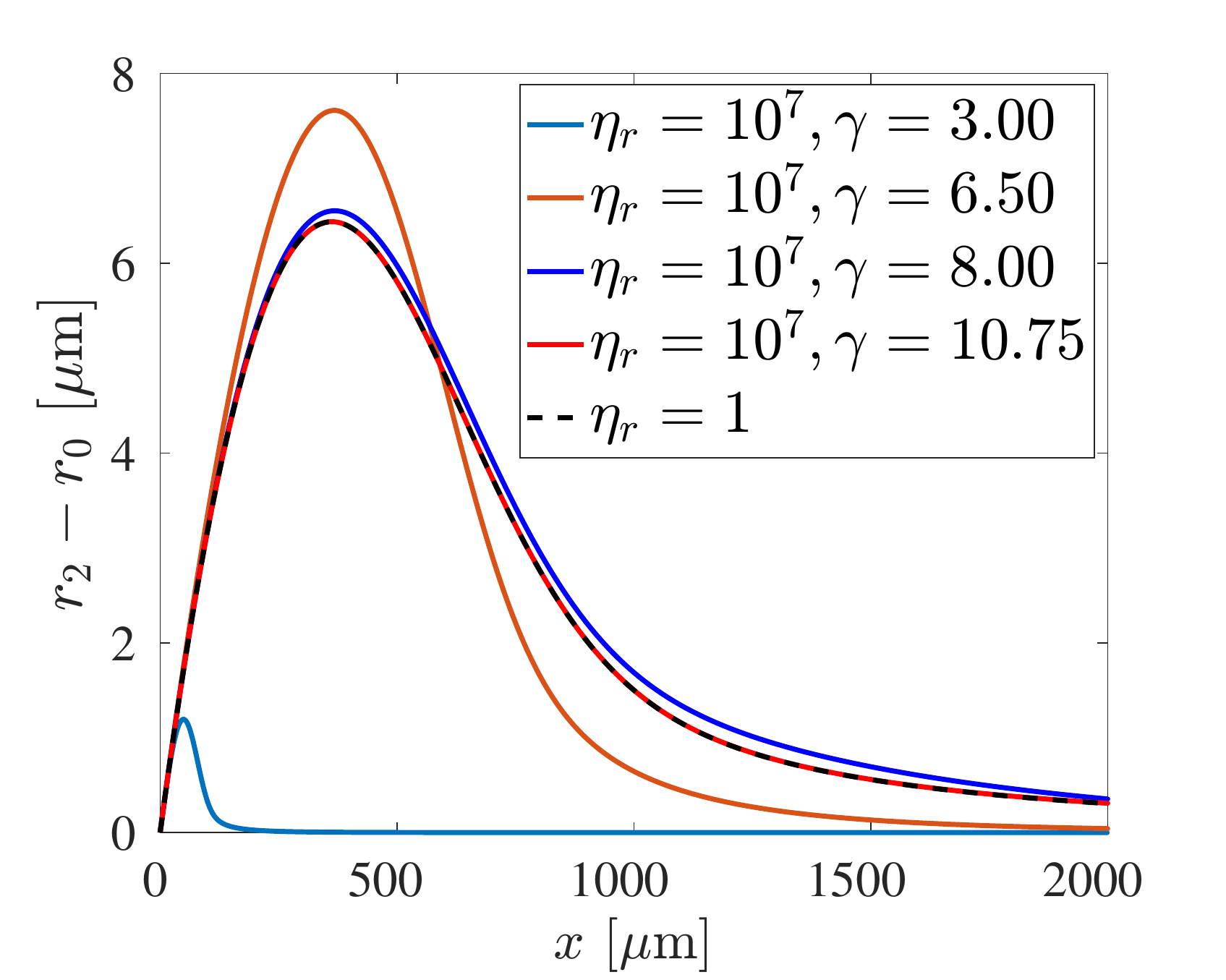} 
\caption{Dimensional deformation as a function of position at dimensional time $10^2$ h for the parameters shown in the inset, and $\mathcal{G} = 10^2$ K/m, $n = 3$, $d_0 = 8$ \AA. Note the difference in the scale of the axes relative to Fig. \ref{fig:etar1e7_d0_8angstrom_n3}(a).} 
\label{fig:etar1e7_d0_8angstrom_n3_largeG}
\end{figure}

\begin{figure}[hbtp]
\centering 
\includegraphics[scale=0.5]{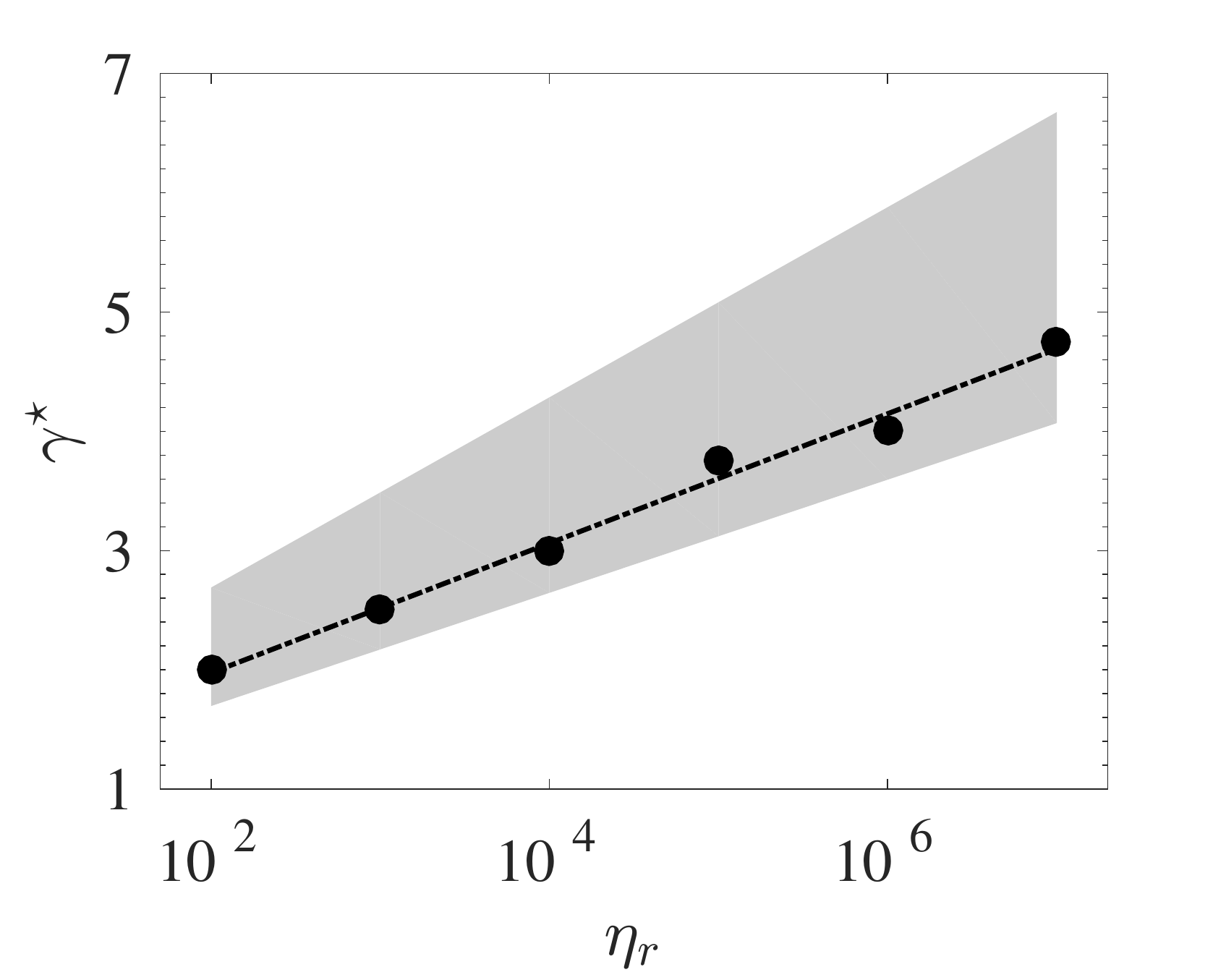} 
\caption{For $n = 3, ~ d_0 = 8$ \AA, $\mathcal{G} = 1$ K/m, we obtain the relation between $\gamma^{\star}$ and $\eta_r$. Symbols are the data points obtained from the numerical solutions. The dash-dotted line represents the linear least square fit: $\gamma^{\star} = 0.8905 + 0.2358 \ln \eta_r$. The shaded region corresponds to the range of $\gamma$ such that $r_{\rm max}(T_f; \eta_r > 1) > r_{\rm max}(T_f; \eta_r = 1)$.} 
\label{fig:gammastar}
\end{figure}

{\bf Similarity solution for} $\boldsymbol{0 < \gamma \leq \gamma_{\rm crit}(\eta_r, d_0, n)}$: We compute the similarity solution, $g_{\rm \gamma}(\zeta_{\rm \gamma})$, numerically by rescaling the deformation, $r(x,t)$, and the position, $x$, using Eqs. \eqref{eq:new_similarity_sol} and \eqref{eq:new_similarity_variable}. Fig. \ref{fig:different_similarity_sol}(a) shows the similarity solution $g_2(\zeta_2)$ for $\eta_r = 10^7$.  

{\bf Breakdown of similarity solution for} $\boldsymbol{ \gamma_{\rm crit}(\eta_r, d_0, n) < \gamma < \gamma_{\rm sat}(\eta_r, d_0, n)}$: For $\eta_r = 10^7$ and $\forall \gamma \in (\gamma_{\rm crit}, \gamma_{\rm sat})$, we compute both $g_{\gamma}(\zeta_{\gamma})$ and $g(\zeta)$ numerically. We find that none of the functions $g(\zeta)$ and $g_{\gamma}(\zeta_{\gamma})$ exhibits self-similarity. For $\eta_r = 10^7$, we plot $g_{4.75}(\zeta_{4.75})$ in Fig. \ref{fig:different_similarity_sol}(b) corresponding to the dimensional time $1$, $10$, and $10^2$ h, where the breakdown of the self-similarity is shown. We have also confirmed that $g(\zeta)$ does not preserve self-similarity (not shown). 

{\bf Similarity solution for} $\boldsymbol{\gamma \geq \gamma_{\rm sat}(\eta_r, d_0, n)}$: As in the case of $0 < \gamma < \gamma_{\rm sat}(\eta_r, d_0, n)$, we compute  $g_{\gamma}(\zeta_{\gamma})$ and $g(\zeta)$ numerically and, as explained in \S \ref{subsubsec:analytical}, only $g(\zeta)$ reflects the self-similar behavior of the deformation (not shown). 

\subsubsection{Capillary Deformation \label{subsubsec:deform}} 

Fig. \ref{fig:etar1e7_d0_8angstrom_n3} shows the dimensional capillary deformation as a function of position at dimensional time $10^2$ h for different values of the power law exponent, $\gamma$, from the three sub-intervals: (i) $(0, \gamma_{\rm crit}]$, (ii) $(\gamma_{\rm crit}, \gamma_{\rm sat})$, and (iii) $[\gamma_{\rm sat}, \infty)$. For $\gamma \in (0, \gamma_{\rm crit})$, the dynamic viscosity of the liquid is dominated by the proximity effect; the liquid volume flux ($\mathcal{Q}$) is reduced relative to that corresponding to the bulk viscosity model ($\mathcal{Q}_b$) across the entire region from the warm to the cold end. Whence, there is less net solidification and capillary deformation relative to the bulk viscosity model \cite{Wettlaufer1995} (c.f, Fig. \ref{fig:etar1e7_d0_8angstrom_n3}(a) versus (b)). This is in accordance with the similarity solution $g_{\gamma}(\zeta_{\gamma})$. 

For $\gamma_{\rm crit} < \gamma < \gamma_{\rm sat}$, the initial solidification and hence capillary deformation in the region $x \leq x_{\rm cross}$ is almost identical for both of the viscosity models, which deviate from each other only in the region $x > x_{\rm cross}$. In other words, 
$\mathcal{Q}(t = 0) = \mathcal{Q}_b(t = 0)$ for $x \leq x_{\rm cross}$ and $\mathcal{Q}(t = 0) < \mathcal{Q}_b(t = 0)$ for $x > x_{\rm cross}$. Because the viscosity at the warm end where the film thickness is large remains unaltered relative to the bulk value, so too is the volume flux at the warm end. Therefore, whilst the thermomolecular pressure gradient depends on the temperature gradient, the gradient in the volume flux depends on the gradient in the film thickness. The latter differs in general between the case with the bulk viscosity \cite{Wettlaufer1995} and that with the proximity effect. However, this difference only manifests itself when the film gets sufficiently thin to distinguish the differences in viscosities and thus the capillary deformation. Hence, the sudden drop in the volume flux at $x \approx x_{\rm cross}$ is accompanied by more solidification in the region $x < x_{\rm cross}$ for any $\gamma \in (\gamma_{\rm crit}, \gamma_{\rm sat})$, when compared to the case of a constant bulk viscosity, or $\gamma \geq \gamma_{\rm sat}$ (see the curves corresponding to $\eta_r = 10^7, ~ \gamma = 4, 4.75, 6.5$ and $\eta_r = 1$ in Fig. \ref{fig:etar1e7_d0_8angstrom_n3}(a)). In this process for some $\gamma$ from the range $(\gamma_{\rm crit}, \gamma_{\rm sat})$, we obtain $r_{\rm max}(T_f; \eta_r \neq 1, \gamma) > r_{\rm max}(T_f; \eta_r \neq 1, \gamma_{\rm sat})$ (or, $r_{\rm max}(T_f; \eta_r = 1)$). 

Independent of the magnitude of the maximum deformation, $r_{\rm max}$, we observe a universal behavior in its position, $x_{\rm max}$. For any $d_0, ~ n, ~ \lambda_n$ and $\eta_r$, we obtain $x_{\rm max}(T_f; \gamma_i) < x_{\rm max}(T_f; \gamma_j)$ for $0< \gamma_i < \gamma_j \leq \gamma_{\rm sat}$. The gradient of the volume flux, $\partial_x \mathcal{Q}$, for $0 < \gamma < \gamma_{\rm sat}$, is larger than that for $\gamma \geq \gamma_{\rm sat}$ (or, $\eta_r = 1$), and this gradient decreases as $\gamma$ increases. In other words, for the same temperature gradient in both the cases, the location of the maximum solidification (deformation of the capillary wall) moves towards the warm end as $\gamma$ decreases.  Finally, as seen when comparing Figs. \ref{fig:etar1e7_d0_8angstrom_n3_largeG} and  \ref{fig:etar1e7_d0_8angstrom_n3}(a), a larger temperature gradient results in a larger deformation that is more highly localized larger temperatures.

\subsubsection{$\gamma^{\star}(\eta_r, d_0, n)$ \label{subsubsec:gammastar}} 

We close this section by computing 

\begin{equation}
\label{eq:gammastar_numerical} 
\gamma^{\star}(\eta_r, d_0, n) = \argmax_{\gamma} \; r_{\rm max}(t = T_f; \gamma, \eta_r, d_0, n), 
\end{equation}
which corresponds to the power law exponent that is associated with the largest maximum deformation of the capillary wall at $T_f$ for given values of $\eta_r, ~ d_0$, and $n$. Fig. \ref{fig:gammastar} shows $\gamma^{\star}$ as a function of $\eta_r$ for $d_0 = 8$ \AA, $n = 3$, and $\mathcal{G} = 1$ K/m. A linear least square fit of the numerically computed $\gamma^{\star}$ yields, $\gamma^{\star} = 0.8905 + 0.2358 \ln \eta_r$. The shaded region corresponds to the range of $\gamma$ such that $r_{\rm max}(T_f; \eta_r > 1) > r_{\rm max}(T_f; \eta_r = 1)$. 

\section{Conclusions \label{sec:conclusions}}

Motivated by a basic question in surface phase transitions, we have developed a phenomenological theory to treat proximity effects in the context of the dynamics of interfacially premelted liquid films.  The principal phenomenon capturing our interest is that as the temperature is reduced and surface and interfacially melted films thin, as described by Eq. \eqref{eq:film_thickness}, a solid like ordering is imprinted upon the liquid eventually spans their thickness.  By parity of reasoning an interface that, upon sufficient increase in temperature, undergoes surface melting is best described as a disordered solid surface, the disorder of which eventually acquires the properties of the bulk liquid.  Whilst we realize that such ordering is not captured by the mean-field theory leading to Eq. \eqref{eq:film_thickness}, that theory nonetheless clearly captures the fact that the thin film is truly a surface phase that would not exist in the absence of the effective field of interactions captured by $\mathcal{I}(d)$ in Eq. \eqref{eq:free_energy}. There are two obvious approaches one might take to address the ordering.  In one, a structurally sensitive surface probe might be employed beginning at very low (high) temperatures and the breakdown (build up) of ordering as the temperature is increased (decreased) along coexistence could be studied. In the other, which we use here, the dynamical consequences of the ordering effects, through their influence on the viscosity of the interfacially melted film play out. In the former case there is little quantitative guidance from experiment and simulations are unable to span the required thermodynamic range \cite[e.g.,][and refs. therein]{Nada:1997, Nada:2000, Nada:2016}.  In the latter case there is ample evidence from a wide range of materials of the effects of confinement on the dynamical properties, although no such measurements have been made {\em in-situ} in a premelted liquid film.  We thus translated the observed proximity effects on the increase of the dynamic viscosity into the setting that has a firm experimental and theoretical foundation for temperatures very near the bulk melting temperature \cite{Wettlaufer1995, Wilen1995, Dash2006, Wettlaufer2006}.  This approach provides an ideal comparative setting.  

Building upon a range of measurements and materials of the increase in the dynamic viscosity upon confinement, we took the simplest approach in Eq. \eqref{eq:powerlaw_viscosity_d} of a power law behavior with film thickness controlled by a single exponent $\gamma$, whose range in several materials is seen in Fig. \ref{fig:viscosity_fit}. The premelted film disjoins its solid phase from a confining elastic capillary tube and the thermomolecular pressure gradient is imposed by setting up a constant temperature gradient parallel to the interface.   In the previous study of this system there is no proximity effect and the film viscosity is the bulk value \cite{Wettlaufer1995}.  The central physical results can be seen through the comparison between the results with and without the proximity effect, as seen in Fig. \ref{fig:etar1e7_d0_8angstrom_n3}(a).   We find that, depending on the strength of the increase in viscosity with change in thickness, determined by $\gamma$, the dynamics of deformation is quantitatively different. For small $\gamma$, the enhanced viscosity is experienced across a larger fraction of the film and as $\gamma$ increases the ordering decays more rapidly and a larger fraction of the film has bulk behavior.  Hence, for small $\gamma$ the viscous resistance to flow is substantial and most of the gradient in the volume flux occurs at the largest temperatures where the film is the thickest.   However, as the gradient in the volume flux controls the capillary deformation, for small $\gamma$ the deformation is correspondingly small. As $\gamma$ increases, so too does the magnitude of the deformation and its position moves away from the entrance to the capillary tube. However, there is a $\gamma$ beyond which the proximity effect saturates and there is no longer a change in the viscosity with film thickness, after which we recover the previous results \cite{Wettlaufer1995} where the capillary tube deforms over a larger region because there is no thickness--and hence temperature--dependence to the viscosity controlling the volume flux.  

Several interesting mathematical results arise in the context of this model.  These principally result from the influence of the viscosity model on the dynamics.   The semi-infinite bulk model admitted similarity solutions \cite{Wettlaufer1995}, but because of the film thickness dependence of Eq. \eqref{eq:powerlaw_viscosity_d} and the temperature dependence of the film thickness, the value of the decay $\gamma$ effectively imposed an external length scale on the problem, which can compromise the similarity behavior, as captured by Fig. \ref{fig:exponent}.  The interplay between the nature of the interactions driving the interfacial melting, captured by $n$, and the viscosity decay, captured by $\gamma$, set the stage for the existence of similarity solutions.  

Finally, the processes discussed here, and in our previous work \cite{Dash2006, Wettlaufer2006}, have implications  for frost heave in a variety of settings, including biological, geological and technological settings, such as the freezing of soils, cells and tissues, cancer treatment, food science, fish biology and botany.  In particular, phase changes and the ordering of water in membranes are active areas of research \cite[e.g.,][]{Parsegian2005}. In particular, our results show that even if the temperature is very low or the pores of a system highly restrictive, the same basic flow driving deformation persists.   So long as there is a temperature gradient, the thermomolecular pressure gradient will eventually build up a deformation near the bulk melting temperature that may be sufficient for rupture of the confining body \cite[e.g.,][and Refs. therein]{Vlahou2010}.  As the proximity effect restricts motion of unfrozen liquid at lower temperatures the deformation will be more highly localized at higher temperatures, as seen by comparing Figs. \ref{fig:etar1e7_d0_8angstrom_n3}(a) and \ref{fig:etar1e7_d0_8angstrom_n3_largeG}. Both the magnitude of the deformation and its position relative to the bulk melting temperature provide experimental targets to indirectly extract constraints, such as the value of $\gamma$, on the proximity effect.  It is hoped that such experiments will be performed. 

\begin{acknowledgements}

The authors acknowledge the support of the Swedish Research Council grant no. 638-2013-9243 and JSW a Royal Society Wolfson Research Merit Award for support.

\end{acknowledgements}

\appendix

\section{Parameters}

\subsection{Parameters for the Ice-Water system \cite{Wettlaufer1996} except for the temperature gradient, which is about 100 times weaker here \label{app:parameters_water}}

\begin{subequations}
\begin{align}
& \mathcal{G} = 1 ~ {\rm K/m}, \label{eq:para1} \\
& k = 3 \times 10^9 ~ {\rm Pa/m}, \label{eq:para2} \\
& \rho_l = 10^3 ~ {\rm kg/m}^{3}, \label{eq:para3} \\
& \rho_s = 0.9167 \times 10^3 ~ {\rm kg/m}^3, \label{eq:para4} \\
& T_m = 273.16 ~ K, \label{eq:para5} \\
& \eta_b = 1.307 \times 10^{-3} ~ {\rm kg/ms}, \label{eq:para7} \\
& \lambda^3 = \Lambda = 1.57 \times 10^{-28} ~ {\rm m}^3 , \label{eq:para8} \\
& q_m = 334 \times 10^3 ~ {\rm J/kg}. \label{eq:para9}
\end{align}
\end{subequations}

\subsubsection{Lists of parameters used in the numerical solutions \label{aubsubsec:tables}}


\begin{table}[!h] 
\begin{tabular}{|c|c|c|}
\hline 
$\gamma$ & $\left(r_2 - r_0\right)_{\rm max}(T_f)$ [$\mu$m] & $x_{\rm max}$ [$\mu$m] \\
\hline 
2.75 & 0.2546 & $1.1250 \times 10^3$ \\ 
3 & 0.2918 & $1.3500 \times 10^3$ \\ 
3.05 & 0.2974 & $1.4000 \times 10^3$ \\
3.1 & 0.3023 & $1.4250 \times 10^3$ \\ 
3.75 & 0.3113 & $1.6750 \times 10^3$ \\ 
4 & 0.3058 & $1.7000 \times 10^3$ \\
4.15 & 0.3034 & $1.7000 \times 10^3$ \\
4.75 & 0.2995 & $1.6750 \times 10^3$ \\ 
5 & 0.2991 & $1.6750 \times 10^3$ \\
5.25 & 0.2989 & $1.6750 \times 10^3$ \\
\hline 
\end{tabular}
\caption{$n = 3, ~ \lambda_n = 5.3947$ \AA, $d_0 = 8$ \AA, $\eta_r = 10^5, ~ \mathcal{G} = 1$ K/m.} 
\end{table}

\begin{table}[!h]
\begin{tabular}{|c|c|c|}
\hline 
$\gamma$ & $\left(r_2 - r_0\right)_{\rm max}(T_f)$ [$\mu$m] & $x_{\rm max}$ [$\mu$m] \\
\hline 
3.25 & 0.2466 & $1.0500 \times 10^3$ \\ 
3.55 & 0.2943 & $1.3000 \times 10^3$ \\ 
3.575 & 0.2975 & $1.3250 \times 10^3$ \\ 
3.625 & 0.3034 & $1.3750 \times 10^3$ \\ 
4 & 0.3241 & $1.6000 \times 10^3$ \\ 
4.25 & 0.3198 & $1.6750 \times 10^3$ \\ 
4.5 & 0.3116 & $1.7000 \times 10^3$ \\ 
4.8 & 0.3046 & $1.7000 \times 10^3$ \\ 
5 & 0.3020 & $1.7000 \times 10^3$ \\ 
5.5 & 0.2994 & $1.6750 \times 10^3$ \\ 
5.75 & 0.2991 & $1.6750 \times 10^3$ \\ 
6 & 0.2989 & $1.6750 \times 10^3$ \\ 
\hline 
\end{tabular}
\caption{$n = 3, ~ \lambda_n = 5.3947$ \AA, $d_0 = 8$ \AA, $\eta_r = 10^6, ~ \mathcal{G} = 1$ K/m.} 
\end{table}

\begin{table}[!h]
\begin{tabular}{|c|c|c|}
\hline 
$\gamma$ & $\left(r_2 - r_0\right)_{\rm max}(T_f)$ [$\mu$m] & $x_{\rm max}$ [$\mu$m] \\
\hline 
0.25 & 0.0023 & $0.1225 \times 10^2$ \\ 
0.33 & 0.0027 & $0.1425 \times 10^2$ \\ 
0.5 & 0.0036 & $0.1875 \times 10^2$ \\ 
0.67 & 0.0048 & $0.2450 \times 10^2$ \\ 
0.75 & 0.0056 & $0.2800 \times 10^2$ \\ 
1 & 0.0084 & $0.4100 \times 10^2$ \\ 
1.5 & 0.0184 & $0.8500 \times 10^2$ \\ 
2 & 0.0361 & $0.1600 \times 10^3$ \\ 
2.5 & 0.0674 & $0.2875 \times 10^3$ \\ 
3 & 0.1184 & $0.5000 \times 10^3$ \\ 
3.5 & 0.1934 & $0.8000 \times 10^3$ \\ 
4 & 0.2805 & $1.1750 \times 10^3$ \\ 
4.1 & 0.2956 & $1.2750 \times 10^3$ \\ 
4.15 & 0.3024 & $1.3000 \times 10^3$ \\ 
4.2 & 0.3086 & $1.3500 \times 10^3$ \\ 
4.75 & 0.3302 & $1.6750 \times 10^3$ \\ 
5 & 0.3204 & $1.7250 \times 10^3$ \\ 
5.325 & 0.3086 & $1.7000 \times 10^3$ \\ 
6 & 0.3000 & $1.6750 \times 10^3$ \\ 
6.5 & 0.2990 & $1.6750 \times 10^3$ \\
6.25 & 0.2993 & $1.6750 \times 10^3$ \\ 
\hline 
\end{tabular}
\caption{$n = 3, ~ \lambda_n = 5.3947$ \AA, $d_0 = 8$ \AA, $\eta_r = 10^7, ~ \mathcal{G} = 1$ K/m.} 
\end{table}

\begin{table}[!h]
\begin{tabular}{|c|c|c|}
\hline 
$\gamma$ & $\left(r_2 - r_0\right)_{\rm max}(T_f)$ [$\mu$m] & $x_{\rm max}$ [$\mu$m] \\
\hline 
3 & 1.1941 & $0.495 \times 10^2$ \\ 
6.5 & 7.6105 & $0.370 \times 10^3$ \\ 
8 & 6.5530 & $0.370 \times 10^3$ \\ 
9 & 6.4494 & $0.365 \times 10^3$ \\ 
9.5 & 6.4411 & $0.365 \times 10^3$ \\ 
10.25 & 6.4380 & $0.360 \times 10^3$ \\ 
10.75 & 6.4375 & $0.360 \times 10^3$ \\
11 & 6.4375 & $0.360 \times 10^3$ \\ 
\hline 
\end{tabular}
\caption{$n = 3, ~ \lambda_n = 5.3947$ \AA, $d_0 = 8$ \AA, $\eta_r = 10^7, ~ \mathcal{G} = 10^2$ K/m.} 
\end{table}

\begin{table}[!h]
\begin{tabular}{|c|c|c|}
\hline 
$\gamma$ & $\left(r_2 - r_0\right)_{\rm max}(T_f)$ [$\mu$m] & $x_{\rm max}$ [$\mu$m] \\
\hline 
0.05 & 0.0239 & $0.110 \times 10^3$ \\ 
0.5 & 0.0598 & $0.250 \times 10^3$ \\ 
1 & 0.1456 & $0.600 \times 10^3$ \\ 
1.5 & 0.3175 & $1.250 \times 10^3$ \\ 
2 & 0.6314 & $2.425 \times 10^3$ \\ 
2.1 & 0.7173 & $2.700 \times 10^3$ \\ 
2.25 & 0.8623 & $3.300 \times 10^3$ \\ 
2.5 & 1.1500 & $4.300 \times 10^3$ \\ 
2.75 & 1.4867 & $5.600 \times 10^3$ \\ 
3 & 1.8372 & $6.975 \times 10^3$ \\ 
3.25 & 2.1221 & $8.400 \times 10^3$ \\ 
3.5 & 2.2547 & $9.400 \times 10^3$ \\ 
3.75 & 2.2459 & $9.900 \times 10^3$ \\ 
4 & 2.1975 & $1.000 \times 10^4$ \\ 
4.25 & 2.1667 & $1.000 \times 10^4$ \\ 
4.5 & 2.1535 & $9.900 \times 10^3$ \\ 
4.75 & 2.1486 & $9.900 \times 10^3$ \\
5 & 2.1469 & $9.900 \times 10^3$ \\ 
\hline 
\end{tabular}
\caption{$n = 2, ~ \lambda_n = 5.3947$ \AA, $d_0 = 8$ \AA, $\eta_r = 10^7, ~ \mathcal{G} = 1$ K/m.} 
\end{table}


\clearpage


\begin{thebibliography}{37}%
\makeatletter
\providecommand \@ifxundefined [1]{%
 \@ifx{#1\undefined}
}%
\providecommand \@ifnum [1]{%
 \ifnum #1\expandafter \@firstoftwo
 \else \expandafter \@secondoftwo
 \fi
}%
\providecommand \@ifx [1]{%
 \ifx #1\expandafter \@firstoftwo
 \else \expandafter \@secondoftwo
 \fi
}%
\providecommand \natexlab [1]{#1}%
\providecommand \enquote  [1]{``#1''}%
\providecommand \bibnamefont  [1]{#1}%
\providecommand \bibfnamefont [1]{#1}%
\providecommand \citenamefont [1]{#1}%
\providecommand \href@noop [0]{\@secondoftwo}%
\providecommand \href [0]{\begingroup \@sanitize@url \@href}%
\providecommand \@href[1]{\@@startlink{#1}\@@href}%
\providecommand \@@href[1]{\endgroup#1\@@endlink}%
\providecommand \@sanitize@url [0]{\catcode `\\12\catcode `\$12\catcode
  `\&12\catcode `\#12\catcode `\^12\catcode `\_12\catcode `\%12\relax}%
\providecommand \@@startlink[1]{}%
\providecommand \@@endlink[0]{}%
\providecommand \url  [0]{\begingroup\@sanitize@url \@url }%
\providecommand \@url [1]{\endgroup\@href {#1}{\urlprefix }}%
\providecommand \urlprefix  [0]{URL }%
\providecommand \Eprint [0]{\href }%
\providecommand \doibase [0]{http://dx.doi.org/}%
\providecommand \selectlanguage [0]{\@gobble}%
\providecommand \bibinfo  [0]{\@secondoftwo}%
\providecommand \bibfield  [0]{\@secondoftwo}%
\providecommand \translation [1]{[#1]}%
\providecommand \BibitemOpen [0]{}%
\providecommand \bibitemStop [0]{}%
\providecommand \bibitemNoStop [0]{.\EOS\space}%
\providecommand \EOS [0]{\spacefactor3000\relax}%
\providecommand \BibitemShut  [1]{\csname bibitem#1\endcsname}%
\let\auto@bib@innerbib\@empty
\bibitem [{\citenamefont {Schick}(1990)}]{SchickLesHouches}%
  \BibitemOpen
  \bibfield  {author} {\bibinfo {author} {\bibfnamefont {M.}~\bibnamefont
  {Schick}},\ }in\ \href@noop {} {\emph {\bibinfo {booktitle} {Les Houches
  Session XLVIII}}}\ (\bibinfo  {publisher} {Elsevier},\ \bibinfo {address}
  {Amsterdam},\ \bibinfo {year} {1990})\ pp.\ \bibinfo {pages}
  {415--497}\BibitemShut {NoStop}%
\bibitem [{\citenamefont {Safran}(1994)}]{Safranbook}%
  \BibitemOpen
  \bibfield  {author} {\bibinfo {author} {\bibfnamefont {S.~A.}\ \bibnamefont
  {Safran}},\ }\href@noop {} {\emph {\bibinfo {title} {Statistical
  Thermodynamics of Surfaces, Interfaces, and Membranes}}}\ (\bibinfo
  {publisher} {Addison-Wesley},\ \bibinfo {address} {Reading, Massachusetts},\
  \bibinfo {year} {1994})\BibitemShut {NoStop}%
\bibitem [{\citenamefont {French}\ \emph {et~al.}(2010)\citenamefont {French},
  \citenamefont {Parsegian}, \citenamefont {Podgornik}, \citenamefont {Rajter},
  \citenamefont {Jagota}, \citenamefont {Luo}, \citenamefont {Asthagiri},
  \citenamefont {Chaudhury}, \citenamefont {Chiang}, \citenamefont {Granick},
  \citenamefont {Kalinin}, \citenamefont {Kardar}, \citenamefont {Kjellander},
  \citenamefont {Langreth}, \citenamefont {Lewis}, \citenamefont {Lustig},
  \citenamefont {Wesolowski}, \citenamefont {Wettlaufer}, \citenamefont
  {Ching}, \citenamefont {Finnis}, \citenamefont {Houlihan}, \citenamefont {von
  Lilienfeld}, \citenamefont {van Oss},\ and\ \citenamefont
  {Zemb}}]{French:2010}%
  \BibitemOpen
  \bibfield  {author} {\bibinfo {author} {\bibfnamefont {R.~H.}\ \bibnamefont
  {French}}, \bibinfo {author} {\bibfnamefont {V.~A.}\ \bibnamefont
  {Parsegian}}, \bibinfo {author} {\bibfnamefont {R.}~\bibnamefont
  {Podgornik}}, \bibinfo {author} {\bibfnamefont {R.~F.}\ \bibnamefont
  {Rajter}}, \bibinfo {author} {\bibfnamefont {A.}~\bibnamefont {Jagota}},
  \bibinfo {author} {\bibfnamefont {J.}~\bibnamefont {Luo}}, \bibinfo {author}
  {\bibfnamefont {D.}~\bibnamefont {Asthagiri}}, \bibinfo {author}
  {\bibfnamefont {M.~K.}\ \bibnamefont {Chaudhury}}, \bibinfo {author}
  {\bibfnamefont {Y.-m.}\ \bibnamefont {Chiang}}, \bibinfo {author}
  {\bibfnamefont {S.}~\bibnamefont {Granick}}, \bibinfo {author} {\bibfnamefont
  {S.}~\bibnamefont {Kalinin}}, \bibinfo {author} {\bibfnamefont
  {M.}~\bibnamefont {Kardar}}, \bibinfo {author} {\bibfnamefont
  {R.}~\bibnamefont {Kjellander}}, \bibinfo {author} {\bibfnamefont {D.~C.}\
  \bibnamefont {Langreth}}, \bibinfo {author} {\bibfnamefont {J.}~\bibnamefont
  {Lewis}}, \bibinfo {author} {\bibfnamefont {S.}~\bibnamefont {Lustig}},
  \bibinfo {author} {\bibfnamefont {D.}~\bibnamefont {Wesolowski}}, \bibinfo
  {author} {\bibfnamefont {J.~S.}\ \bibnamefont {Wettlaufer}}, \bibinfo
  {author} {\bibfnamefont {W.-Y.}\ \bibnamefont {Ching}}, \bibinfo {author}
  {\bibfnamefont {M.}~\bibnamefont {Finnis}}, \bibinfo {author} {\bibfnamefont
  {F.}~\bibnamefont {Houlihan}}, \bibinfo {author} {\bibfnamefont {O.~A.}\
  \bibnamefont {von Lilienfeld}}, \bibinfo {author} {\bibfnamefont {C.~J.}\
  \bibnamefont {van Oss}}, \ and\ \bibinfo {author} {\bibfnamefont
  {T.}~\bibnamefont {Zemb}},\ }\href {\doibase 10.1103/RevModPhys.82.1887}
  {\bibfield  {journal} {\bibinfo  {journal} {Rev. Mod. Phys.}\ }\textbf
  {\bibinfo {volume} {82}},\ \bibinfo {pages} {1887} (\bibinfo {year}
  {2010})}\BibitemShut {NoStop}%
\bibitem [{\citenamefont {Israelachvili}(2011)}]{Jacobbook}%
  \BibitemOpen
  \bibfield  {author} {\bibinfo {author} {\bibfnamefont {J.~N.}\ \bibnamefont
  {Israelachvili}},\ }\href@noop {} {\emph {\bibinfo {title} {Intermolecular
  and Surface Forces}}},\ \bibinfo {edition} {3rd}\ ed.\ (\bibinfo  {publisher}
  {Academic Press},\ \bibinfo {address} {New York, NY},\ \bibinfo {year}
  {2011})\BibitemShut {NoStop}%
\bibitem [{\citenamefont {Dash}\ \emph {et~al.}(2006)\citenamefont {Dash},
  \citenamefont {Rempel},\ and\ \citenamefont {Wettlaufer}}]{Dash2006}%
  \BibitemOpen
  \bibfield  {author} {\bibinfo {author} {\bibfnamefont {J.~G.}\ \bibnamefont
  {Dash}}, \bibinfo {author} {\bibfnamefont {A.~W.}\ \bibnamefont {Rempel}}, \
  and\ \bibinfo {author} {\bibfnamefont {J.~S.}\ \bibnamefont {Wettlaufer}},\
  }\href@noop {} {\bibfield  {journal} {\bibinfo  {journal} {Rev. Mod. Phys.}\
  }\textbf {\bibinfo {volume} {78}} (\bibinfo {year} {2006})}\BibitemShut
  {NoStop}%
\bibitem [{\citenamefont {Wilen}\ \emph {et~al.}(1995)\citenamefont {Wilen},
  \citenamefont {Wettlaufer}, \citenamefont {Elbaum},\ and\ \citenamefont
  {Schick}}]{Wilenetal1995}%
  \BibitemOpen
  \bibfield  {author} {\bibinfo {author} {\bibfnamefont {L.~A.}\ \bibnamefont
  {Wilen}}, \bibinfo {author} {\bibfnamefont {J.~S.}\ \bibnamefont
  {Wettlaufer}}, \bibinfo {author} {\bibfnamefont {M.}~\bibnamefont {Elbaum}},
  \ and\ \bibinfo {author} {\bibfnamefont {M.}~\bibnamefont {Schick}},\
  }\href@noop {} {\bibfield  {journal} {\bibinfo  {journal} {Phys. Rev. B}\
  }\textbf {\bibinfo {volume} {52}},\ \bibinfo {pages} {12,426} (\bibinfo
  {year} {1995})}\BibitemShut {NoStop}%
\bibitem [{\citenamefont {Wettlaufer}\ and\ \citenamefont
  {Worster}(1995)}]{Wettlaufer1995}%
  \BibitemOpen
  \bibfield  {author} {\bibinfo {author} {\bibfnamefont {J.~S.}\ \bibnamefont
  {Wettlaufer}}\ and\ \bibinfo {author} {\bibfnamefont {M.~G.}\ \bibnamefont
  {Worster}},\ }\href {\doibase 10.1103/PhysRevE.51.4679} {\bibfield  {journal}
  {\bibinfo  {journal} {Phys. Rev. E}\ }\textbf {\bibinfo {volume} {51}},\
  \bibinfo {pages} {4679} (\bibinfo {year} {1995})}\BibitemShut {NoStop}%
\bibitem [{\citenamefont {Wettlaufer}\ \emph {et~al.}(1996)\citenamefont
  {Wettlaufer}, \citenamefont {Worster}, \citenamefont {Wilen},\ and\
  \citenamefont {Dash}}]{Wettlaufer1996}%
  \BibitemOpen
  \bibfield  {author} {\bibinfo {author} {\bibfnamefont {J.~S.}\ \bibnamefont
  {Wettlaufer}}, \bibinfo {author} {\bibfnamefont {M.~G.}\ \bibnamefont
  {Worster}}, \bibinfo {author} {\bibfnamefont {L.~A.}\ \bibnamefont {Wilen}},
  \ and\ \bibinfo {author} {\bibfnamefont {J.~G.}\ \bibnamefont {Dash}},\
  }\href {\doibase 10.1103/PhysRevLett.76.3602} {\bibfield  {journal} {\bibinfo
   {journal} {Phys. Rev. Lett.}\ }\textbf {\bibinfo {volume} {76}},\ \bibinfo
  {pages} {3602} (\bibinfo {year} {1996})}\BibitemShut {NoStop}%
\bibitem [{\citenamefont {de~Gennes}(1985)}]{deGennes}%
  \BibitemOpen
  \bibfield  {author} {\bibinfo {author} {\bibfnamefont {P.~G.}\ \bibnamefont
  {de~Gennes}},\ }\href@noop {} {\bibfield  {journal} {\bibinfo  {journal}
  {Rev. Mod. Phys.}\ }\textbf {\bibinfo {volume} {57}} (\bibinfo {year}
  {1985})}\BibitemShut {NoStop}%
\bibitem [{\citenamefont {Wettlaufer}\ and\ \citenamefont
  {Worster}(2006)}]{Wettlaufer2006}%
  \BibitemOpen
  \bibfield  {author} {\bibinfo {author} {\bibfnamefont {J.~S.}\ \bibnamefont
  {Wettlaufer}}\ and\ \bibinfo {author} {\bibfnamefont {M.~G.}\ \bibnamefont
  {Worster}},\ }\href@noop {} {\bibfield  {journal} {\bibinfo  {journal} {Annu.
  Rev. Fl. Mech.}\ }\textbf {\bibinfo {volume} {38}},\ \bibinfo {pages} {427}
  (\bibinfo {year} {2006})}\BibitemShut {NoStop}%
\bibitem [{\citenamefont {Wilen}\ and\ \citenamefont {Dash}(1995)}]{Wilen1995}%
  \BibitemOpen
  \bibfield  {author} {\bibinfo {author} {\bibfnamefont {L.~A.}\ \bibnamefont
  {Wilen}}\ and\ \bibinfo {author} {\bibfnamefont {J.~G.}\ \bibnamefont
  {Dash}},\ }\href {\doibase 10.1103/PhysRevLett.74.5076} {\bibfield  {journal}
  {\bibinfo  {journal} {Phys. Rev. Lett.}\ }\textbf {\bibinfo {volume} {74}},\
  \bibinfo {pages} {5076} (\bibinfo {year} {1995})}\BibitemShut {NoStop}%
\bibitem [{\citenamefont {Rempel}\ \emph {et~al.}(2004)\citenamefont {Rempel},
  \citenamefont {Wettlaufer},\ and\ \citenamefont {Worster}}]{Rempel2004}%
  \BibitemOpen
  \bibfield  {author} {\bibinfo {author} {\bibfnamefont {A.~W.}\ \bibnamefont
  {Rempel}}, \bibinfo {author} {\bibfnamefont {J.~S.}\ \bibnamefont
  {Wettlaufer}}, \ and\ \bibinfo {author} {\bibfnamefont {M.~G.}\ \bibnamefont
  {Worster}},\ }\href@noop {} {\bibfield  {journal} {\bibinfo  {journal} {J.
  Fluid Mech.}\ }\textbf {\bibinfo {volume} {498}},\ \bibinfo {pages} {227}
  (\bibinfo {year} {2004})}\BibitemShut {NoStop}%
\bibitem [{\citenamefont {Zhu}\ \emph {et~al.}(2000)\citenamefont {Zhu},
  \citenamefont {Vilches}, \citenamefont {Dash}, \citenamefont {Sing},\ and\
  \citenamefont {Wettlaufer}}]{Zhu2000}%
  \BibitemOpen
  \bibfield  {author} {\bibinfo {author} {\bibfnamefont {D.-M.}\ \bibnamefont
  {Zhu}}, \bibinfo {author} {\bibfnamefont {O.~E.}\ \bibnamefont {Vilches}},
  \bibinfo {author} {\bibfnamefont {J.~G.}\ \bibnamefont {Dash}}, \bibinfo
  {author} {\bibfnamefont {B.}~\bibnamefont {Sing}}, \ and\ \bibinfo {author}
  {\bibfnamefont {J.~S.}\ \bibnamefont {Wettlaufer}},\ }\href {\doibase
  10.1103/PhysRevLett.85.4908} {\bibfield  {journal} {\bibinfo  {journal}
  {Phys. Rev. Lett.}\ }\textbf {\bibinfo {volume} {85}},\ \bibinfo {pages}
  {4908} (\bibinfo {year} {2000})}\BibitemShut {NoStop}%
\bibitem [{\citenamefont {Taber}(1929)}]{Taber1929}%
  \BibitemOpen
  \bibfield  {author} {\bibinfo {author} {\bibfnamefont {S.}~\bibnamefont
  {Taber}},\ }\href@noop {} {\bibfield  {journal} {\bibinfo  {journal} {J.
  Geol.}\ }\textbf {\bibinfo {volume} {37}},\ \bibinfo {pages} {428} (\bibinfo
  {year} {1929})}\BibitemShut {NoStop}%
\bibitem [{\citenamefont {Mizusaki}\ and\ \citenamefont
  {Hiroi}(1995)}]{Mizusaki1995}%
  \BibitemOpen
  \bibfield  {author} {\bibinfo {author} {\bibfnamefont {T.}~\bibnamefont
  {Mizusaki}}\ and\ \bibinfo {author} {\bibfnamefont {M.}~\bibnamefont
  {Hiroi}},\ }\href@noop {} {\bibfield  {journal} {\bibinfo  {journal} {Physica
  B}\ }\textbf {\bibinfo {volume} {210}},\ \bibinfo {pages} {403} (\bibinfo
  {year} {1995})}\BibitemShut {NoStop}%
\bibitem [{\citenamefont {Evans}(1990)}]{Evans1990}%
  \BibitemOpen
  \bibfield  {author} {\bibinfo {author} {\bibfnamefont {R.}~\bibnamefont
  {Evans}},\ }\href@noop {} {\bibfield  {journal} {\bibinfo  {journal} {J.
  Phys. Cond. Matt.}\ }\textbf {\bibinfo {volume} {2}},\ \bibinfo {pages}
  {8989} (\bibinfo {year} {1990})}\BibitemShut {NoStop}%
\bibitem [{\citenamefont {Christenson}(2001)}]{Hugo2001}%
  \BibitemOpen
  \bibfield  {author} {\bibinfo {author} {\bibfnamefont {H.~K.}\ \bibnamefont
  {Christenson}},\ }\href@noop {} {\bibfield  {journal} {\bibinfo  {journal}
  {J. Phys. Cond. Matt.}\ }\textbf {\bibinfo {volume} {13}},\ \bibinfo {pages}
  {R95} (\bibinfo {year} {2001})}\BibitemShut {NoStop}%
\bibitem [{\citenamefont {Koplik}\ and\ \citenamefont
  {Banavar}(1995)}]{Koplik1995}%
  \BibitemOpen
  \bibfield  {author} {\bibinfo {author} {\bibfnamefont {J.}~\bibnamefont
  {Koplik}}\ and\ \bibinfo {author} {\bibfnamefont {J.~R.}\ \bibnamefont
  {Banavar}},\ }\href@noop {} {\bibfield  {journal} {\bibinfo  {journal} {Annu.
  Rev. Fl. Mech.}\ }\textbf {\bibinfo {volume} {27}},\ \bibinfo {pages} {257}
  (\bibinfo {year} {1995})}\BibitemShut {NoStop}%
\bibitem [{\citenamefont {Nada}\ and\ \citenamefont
  {Furukawa}(1997)}]{Nada:1997}%
  \BibitemOpen
  \bibfield  {author} {\bibinfo {author} {\bibfnamefont {H.}~\bibnamefont
  {Nada}}\ and\ \bibinfo {author} {\bibfnamefont {Y.}~\bibnamefont
  {Furukawa}},\ }\href@noop {} {\bibfield  {journal} {\bibinfo  {journal} {J.
  Phys. Chem. B}\ }\textbf {\bibinfo {volume} {101}},\ \bibinfo {pages} {6163}
  (\bibinfo {year} {1997})}\BibitemShut {NoStop}%
\bibitem [{\citenamefont {Nada}\ and\ \citenamefont
  {Furukawa}(2000)}]{Nada:2000}%
  \BibitemOpen
  \bibfield  {author} {\bibinfo {author} {\bibfnamefont {H.}~\bibnamefont
  {Nada}}\ and\ \bibinfo {author} {\bibfnamefont {Y.}~\bibnamefont
  {Furukawa}},\ }\href@noop {} {\bibfield  {journal} {\bibinfo  {journal}
  {Surf. Sci.}\ }\textbf {\bibinfo {volume} {446}},\ \bibinfo {pages} {1}
  (\bibinfo {year} {2000})}\BibitemShut {NoStop}%
\bibitem [{\citenamefont {Nada}(2016)}]{Nada:2016}%
  \BibitemOpen
  \bibfield  {author} {\bibinfo {author} {\bibfnamefont {H.}~\bibnamefont
  {Nada}},\ }\href {\doibase 10.1063/1.4973000} {\bibfield  {journal} {\bibinfo
   {journal} {J. Chem. Phys.}\ }\textbf {\bibinfo {volume} {145}} (\bibinfo
  {year} {2016}),\ 10.1063/1.4973000}\BibitemShut {NoStop}%
\bibitem [{\citenamefont {Pittenger}\ \emph {et~al.}(2001)\citenamefont
  {Pittenger}, \citenamefont {Fain}, \citenamefont {Cochran}, \citenamefont
  {Donev}, \citenamefont {Robertson}, \citenamefont {Szuchmacher},\ and\
  \citenamefont {Overney}}]{Pittenger2001}%
  \BibitemOpen
  \bibfield  {author} {\bibinfo {author} {\bibfnamefont {B.}~\bibnamefont
  {Pittenger}}, \bibinfo {author} {\bibfnamefont {S.~C.}\ \bibnamefont {Fain}},
  \bibinfo {author} {\bibfnamefont {M.~J.}\ \bibnamefont {Cochran}}, \bibinfo
  {author} {\bibfnamefont {J.~M.~K.}\ \bibnamefont {Donev}}, \bibinfo {author}
  {\bibfnamefont {B.~E.}\ \bibnamefont {Robertson}}, \bibinfo {author}
  {\bibfnamefont {A.}~\bibnamefont {Szuchmacher}}, \ and\ \bibinfo {author}
  {\bibfnamefont {R.~M.}\ \bibnamefont {Overney}},\ }\href {\doibase
  10.1103/PhysRevB.63.134102} {\bibfield  {journal} {\bibinfo  {journal} {Phys.
  Rev. B}\ }\textbf {\bibinfo {volume} {63}},\ \bibinfo {pages} {134102}
  (\bibinfo {year} {2001})}\BibitemShut {NoStop}%
\bibitem [{\citenamefont {D\"{o}ppenschmidt}\ \emph {et~al.}(1998)\citenamefont
  {D\"{o}ppenschmidt}, \citenamefont {Kappl},\ and\ \citenamefont
  {Butt}}]{Butt1998}%
  \BibitemOpen
  \bibfield  {author} {\bibinfo {author} {\bibfnamefont {A.}~\bibnamefont
  {D\"{o}ppenschmidt}}, \bibinfo {author} {\bibfnamefont {M.}~\bibnamefont
  {Kappl}}, \ and\ \bibinfo {author} {\bibfnamefont {H.-J.}\ \bibnamefont
  {Butt}},\ }\href@noop {} {\bibfield  {journal} {\bibinfo  {journal} {J. Phys.
  Chem. B}\ }\textbf {\bibinfo {volume} {102}},\ \bibinfo {pages} {7813}
  (\bibinfo {year} {1998})}\BibitemShut {NoStop}%
\bibitem [{\citenamefont {Dash}(2003)}]{Dash2003}%
  \BibitemOpen
  \bibfield  {author} {\bibinfo {author} {\bibfnamefont {J.~G.}\ \bibnamefont
  {Dash}},\ }\href@noop {} {\bibfield  {journal} {\bibinfo  {journal} {Scripta
  Mat.}\ }\textbf {\bibinfo {volume} {49}},\ \bibinfo {pages} {1003} (\bibinfo
  {year} {2003})}\BibitemShut {NoStop}%
\bibitem [{\citenamefont {Dhinojwala}\ and\ \citenamefont
  {Granick}(1997)}]{Dhinojwala1997}%
  \BibitemOpen
  \bibfield  {author} {\bibinfo {author} {\bibfnamefont {A.}~\bibnamefont
  {Dhinojwala}}\ and\ \bibinfo {author} {\bibfnamefont {S.}~\bibnamefont
  {Granick}},\ }\href {\doibase 10.1021/ja9632318} {\bibfield  {journal}
  {\bibinfo  {journal} {J. Am. Chem. Soc.}\ }\textbf {\bibinfo {volume}
  {119}},\ \bibinfo {pages} {241} (\bibinfo {year} {1997})}\BibitemShut
  {NoStop}%
\bibitem [{\citenamefont {Major}\ \emph {et~al.}(2006)\citenamefont {Major},
  \citenamefont {Houston}, \citenamefont {McGrath}, \citenamefont {Siepmann},\
  and\ \citenamefont {Zhu}}]{Major2006}%
  \BibitemOpen
  \bibfield  {author} {\bibinfo {author} {\bibfnamefont {R.~C.}\ \bibnamefont
  {Major}}, \bibinfo {author} {\bibfnamefont {J.~E.}\ \bibnamefont {Houston}},
  \bibinfo {author} {\bibfnamefont {M.~J.}\ \bibnamefont {McGrath}}, \bibinfo
  {author} {\bibfnamefont {J.~I.}\ \bibnamefont {Siepmann}}, \ and\ \bibinfo
  {author} {\bibfnamefont {X.-Y.}\ \bibnamefont {Zhu}},\ }\href {\doibase
  10.1103/PhysRevLett.96.177803} {\bibfield  {journal} {\bibinfo  {journal}
  {Phys. Rev. Lett.}\ }\textbf {\bibinfo {volume} {96}},\ \bibinfo {pages}
  {177803} (\bibinfo {year} {2006})}\BibitemShut {NoStop}%
\bibitem [{\citenamefont {Bureau}(2010)}]{Bureau2010}%
  \BibitemOpen
  \bibfield  {author} {\bibinfo {author} {\bibfnamefont {L.}~\bibnamefont
  {Bureau}},\ }\href {\doibase 10.1103/PhysRevLett.104.218302} {\bibfield
  {journal} {\bibinfo  {journal} {Phys. Rev. Lett.}\ }\textbf {\bibinfo
  {volume} {104}},\ \bibinfo {pages} {218302} (\bibinfo {year}
  {2010})}\BibitemShut {NoStop}%
\bibitem [{\citenamefont {Raviv}\ and\ \citenamefont
  {Klein}(2002)}]{Raviv2002}%
  \BibitemOpen
  \bibfield  {author} {\bibinfo {author} {\bibfnamefont {U.}~\bibnamefont
  {Raviv}}\ and\ \bibinfo {author} {\bibfnamefont {J.}~\bibnamefont {Klein}},\
  }\href {\doibase 10.1126/science.1074481} {\bibfield  {journal} {\bibinfo
  {journal} {Science}\ }\textbf {\bibinfo {volume} {297}},\ \bibinfo {pages}
  {1540} (\bibinfo {year} {2002})}\BibitemShut {NoStop}%
\bibitem [{\citenamefont {Granick}(1991)}]{Granick1991}%
  \BibitemOpen
  \bibfield  {author} {\bibinfo {author} {\bibfnamefont {S.}~\bibnamefont
  {Granick}},\ }\href {http://www.jstor.org/stable/2878740} {\bibfield
  {journal} {\bibinfo  {journal} {Science}\ }\textbf {\bibinfo {volume}
  {253}},\ \bibinfo {pages} {1374} (\bibinfo {year} {1991})}\BibitemShut
  {NoStop}%
\bibitem [{\citenamefont {Zhu}\ and\ \citenamefont {Granick}(2003)}]{Zhu2003}%
  \BibitemOpen
  \bibfield  {author} {\bibinfo {author} {\bibfnamefont {Y.}~\bibnamefont
  {Zhu}}\ and\ \bibinfo {author} {\bibfnamefont {S.}~\bibnamefont {Granick}},\
  }\href {\doibase 10.1021/la035155+} {\bibfield  {journal} {\bibinfo
  {journal} {Langmuir}\ }\textbf {\bibinfo {volume} {19}},\ \bibinfo {pages}
  {8148} (\bibinfo {year} {2003})}\BibitemShut {NoStop}%
\bibitem [{\citenamefont {Zhu}\ and\ \citenamefont {Granick}(2004)}]{Zhu2004}%
  \BibitemOpen
  \bibfield  {author} {\bibinfo {author} {\bibfnamefont {Y.}~\bibnamefont
  {Zhu}}\ and\ \bibinfo {author} {\bibfnamefont {S.}~\bibnamefont {Granick}},\
  }\href@noop {} {\bibfield  {journal} {\bibinfo  {journal} {Phys. Rev. Lett.}\
  }\textbf {\bibinfo {volume} {93}},\ \bibinfo {pages} {096101} (\bibinfo
  {year} {2004})}\BibitemShut {NoStop}%
\bibitem [{\citenamefont {Becker}\ and\ \citenamefont
  {Mugele}(2003)}]{Becker2003}%
  \BibitemOpen
  \bibfield  {author} {\bibinfo {author} {\bibfnamefont {T.}~\bibnamefont
  {Becker}}\ and\ \bibinfo {author} {\bibfnamefont {F.}~\bibnamefont
  {Mugele}},\ }\href {\doibase 10.1103/PhysRevLett.91.166104} {\bibfield
  {journal} {\bibinfo  {journal} {Phys. Rev. Lett.}\ }\textbf {\bibinfo
  {volume} {91}},\ \bibinfo {pages} {166104} (\bibinfo {year}
  {2003})}\BibitemShut {NoStop}%
\bibitem [{\citenamefont {LeVeque}(2007)}]{LeVeque2007a}%
  \BibitemOpen
  \bibfield  {author} {\bibinfo {author} {\bibfnamefont {R.}~\bibnamefont
  {LeVeque}},\ }\href {\doibase 10.1137/1.9780898717839} {\emph {\bibinfo
  {title} {Finite Difference Methods for Ordinary and Partial Differential
  Equations}}}\ (\bibinfo  {publisher} {Society for Industrial and Applied
  Mathematics},\ \bibinfo {year} {2007})\BibitemShut {NoStop}%
\bibitem [{\citenamefont {Toppaladoddi}\ and\ \citenamefont
  {Wettlaufer}(2017)}]{Toppaladoddi2017}%
  \BibitemOpen
  \bibfield  {author} {\bibinfo {author} {\bibfnamefont {S.}~\bibnamefont
  {Toppaladoddi}}\ and\ \bibinfo {author} {\bibfnamefont {J.~S.}\ \bibnamefont
  {Wettlaufer}},\ }\href {\doibase 10.1007/s10955-016-1704-8} {\bibfield
  {journal} {\bibinfo  {journal} {J. Stat. Phys.}\ }\textbf {\bibinfo {volume}
  {167}},\ \bibinfo {pages} {683} (\bibinfo {year} {2017})}\BibitemShut
  {NoStop}%
\bibitem [{SM()}]{SM}%
  \BibitemOpen
  \href@noop {} {}\bibinfo {note} {See Supplemental Material for the numerical
  scheme and additional range of $\gamma$ explored, but not shown
  here.}\BibitemShut {Stop}%
\bibitem [{\citenamefont {Parsegian}(2005)}]{Parsegian2005}%
  \BibitemOpen
  \bibfield  {author} {\bibinfo {author} {\bibfnamefont {V.~A.}\ \bibnamefont
  {Parsegian}},\ }\href {\doibase 10.1017/CBO9780511614606} {\emph {\bibinfo
  {title} {Van der Waals Forces: A Handbook for Biologists, Chemists,
  Engineers, and Physicists}}}\ (\bibinfo  {publisher} {Cambridge University
  Press},\ \bibinfo {year} {2005})\BibitemShut {NoStop}%
\bibitem [{\citenamefont {Vlahou}\ and\ \citenamefont
  {Worster}(2010)}]{Vlahou2010}%
  \BibitemOpen
  \bibfield  {author} {\bibinfo {author} {\bibfnamefont {I.}~\bibnamefont
  {Vlahou}}\ and\ \bibinfo {author} {\bibfnamefont {M.~G.}\ \bibnamefont
  {Worster}},\ }\href@noop {} {\bibfield  {journal} {\bibinfo  {journal} {J.
  Glaciol.}\ }\textbf {\bibinfo {volume} {56}},\ \bibinfo {pages} {271}
  (\bibinfo {year} {2010})}\BibitemShut {NoStop}%
\end{thebibliography}

\begin{thebibliography}{2}%
\makeatletter
\providecommand \@ifxundefined [1]{%
 \@ifx{#1\undefined}
}%
\providecommand \@ifnum [1]{%
 \ifnum #1\expandafter \@firstoftwo
 \else \expandafter \@secondoftwo
 \fi
}%
\providecommand \@ifx [1]{%
 \ifx #1\expandafter \@firstoftwo
 \else \expandafter \@secondoftwo
 \fi
}%
\providecommand \natexlab [1]{#1}%
\providecommand \enquote  [1]{``#1''}%
\providecommand \bibnamefont  [1]{#1}%
\providecommand \bibfnamefont [1]{#1}%
\providecommand \citenamefont [1]{#1}%
\providecommand \href@noop [0]{\@secondoftwo}%
\providecommand \href [0]{\begingroup \@sanitize@url \@href}%
\providecommand \@href[1]{\@@startlink{#1}\@@href}%
\providecommand \@@href[1]{\endgroup#1\@@endlink}%
\providecommand \@sanitize@url [0]{\catcode `\\12\catcode `\$12\catcode
  `\&12\catcode `\#12\catcode `\^12\catcode `\_12\catcode `\%12\relax}%
\providecommand \@@startlink[1]{}%
\providecommand \@@endlink[0]{}%
\providecommand \url  [0]{\begingroup\@sanitize@url \@url }%
\providecommand \@url [1]{\endgroup\@href {#1}{\urlprefix }}%
\providecommand \urlprefix  [0]{URL }%
\providecommand \Eprint [0]{\href }%
\providecommand \doibase [0]{http://dx.doi.org/}%
\providecommand \selectlanguage [0]{\@gobble}%
\providecommand \bibinfo  [0]{\@secondoftwo}%
\providecommand \bibfield  [0]{\@secondoftwo}%
\providecommand \translation [1]{[#1]}%
\providecommand \BibitemOpen [0]{}%
\providecommand \bibitemStop [0]{}%
\providecommand \bibitemNoStop [0]{.\EOS\space}%
\providecommand \EOS [0]{\spacefactor3000\relax}%
\providecommand \BibitemShut  [1]{\csname bibitem#1\endcsname}%
\let\auto@bib@innerbib\@empty
\bibitem [{\citenamefont {LeVeque}(2007)}]{LeVeque2007SM}%
  \BibitemOpen
  \bibfield  {author} {\bibinfo {author} {\bibfnamefont {R.}~\bibnamefont
  {LeVeque}},\ }\href {\doibase 10.1137/1.9780898717839} {\emph {\bibinfo
  {title} {Finite Difference Methods for Ordinary and Partial Differential
  Equations}}}\ (\bibinfo  {publisher} {Society for Industrial and Applied
  Mathematics},\ \bibinfo {year} {2007})\BibitemShut {NoStop}%
\bibitem [{\citenamefont {Toppaladoddi}\ and\ \citenamefont
  {Wettlaufer}(2017)}]{Toppaladoddi2017SM}%
  \BibitemOpen
  \bibfield  {author} {\bibinfo {author} {\bibfnamefont {S.}~\bibnamefont
  {Toppaladoddi}}\ and\ \bibinfo {author} {\bibfnamefont {J.~S.}\ \bibnamefont
  {Wettlaufer}},\ }\href {\doibase 10.1007/s10955-016-1704-8} {\bibfield
  {journal} {\bibinfo  {journal} {J. Stat. Phys.}\ }\textbf {\bibinfo {volume}
  {167}},\ \bibinfo {pages} {683} (\bibinfo {year} {2017})}\BibitemShut
  {NoStop}%
\end{thebibliography}


%

\clearpage
\widetext
\begin{center}
\textbf{\large Confinement effects in premelting dynamics: Supplemental Material}
\end{center}
\setcounter{equation}{0}
\setcounter{figure}{0}
\setcounter{table}{0}
\setcounter{section}{0}
\makeatletter
\renewcommand{\theequation}{S\arabic{equation}}
\renewcommand{\thefigure}{S\arabic{figure}}
\renewcommand{\thetable}{S\arabic{table}}
\renewcommand{\bibnumfmt}[1]{[S#1]}
\renewcommand{\citenumfont}[1]{S#1}

\begin{widetext}

\begin{center}
\textbf{\uppercase{\small S-I. Comment on the derivation of Eq. (11) of the main text}}
\end{center}

From Eqs. (7)-(9) \& (14) we can write 

\begin{eqnarray}
& & \partial_t r_2 - \Gamma_3 \frac{1}{r_2} \partial_x \left[ \frac{r_2}{x} \left\{ \partial_x \left( r_2 - r_0 \right) - \alpha \right\} \right] = 0 \nonumber \\ 
\label{eq:mass-conservation} 
& & \partial_t r_2 - \Gamma_3 \underbrace{ \partial_x \left[ \frac{ \partial_x \left( r_2 - r_0 \right) - \alpha }{ x } \right] }_{\rm A} - \Gamma_3 \underbrace{ \frac{ \partial_x \left( r_2 - r_0 \right) - \alpha }{ x } \frac{ \partial_x r_2 }{ r_2 } }_{\rm B} = 0. 
\end{eqnarray}
Now, let $\mathcal{R}(x,t) \equiv \frac{ \partial_x \left( r_2 - r_0 \right) - \alpha }{ x }$ and consider the magnitude of the ratio of the second and third term in \eqref{eq:mass-conservation} as

\begin{eqnarray}
\frac{\rm A}{\rm B} 
& = & \frac{ \partial_x \ln\mathcal{R}(x,t)}{\frac{\partial_x r_2 }{ r_2 } }, 
\end{eqnarray}
in which the key issue is what happens at small $x$, wherein the boundary and necessary conditions Eqs. 12(a-c) are important. Hence, we examine the small $x$ limit of ${\rm A}/{\rm B}$.  First examine the numerator viz., 

\begin{equation}
\lim_{x\to0} \mathcal{R}(x,t) = \partial_x [\partial_x r_2] = 0, \implies \lim_{x\to0} |\ln\mathcal{R}(x,t)| \rightarrow \infty, 
\end{equation}
whereas 

\begin{equation}
\lim_{x\to0} B = \lim_{x\to0} {\frac{\partial_x r_2 }{ r_2 } } = {\frac{\alpha}{ r_0 } } , 
\end{equation}
Therefore, near the warm end, where $x$ is small and most of the deformation occurs, we have ${\rm A} \gg {\rm B}$. Moreover, at the nearby local maximum, $\mathcal{R}(x,t) \propto x^{-1}$ whereas ${\partial_x r_2 } = 0$ so that ${\rm B}$ vanishes. Whence, we neglect term ${\rm B}$ in Eq. \eqref{eq:mass-conservation} and obtain 

\begin{equation}
\label{eq:final_mass_conservation}
\partial_t r_2 - \Gamma_3 \partial_x \left[ \frac{ \partial_x \left( r_2 - r_0 \right) - \alpha }{ x } \right] = 0, 
\end{equation}
which is identical to Eq. (11) for $n = 3$ and $\eta(x) = \eta_b$. As was shown in [7], Eq. (\ref{eq:final_mass_conservation}) is consistent with experiments in the high temperature region.  In the present work we are focused on the confinement effects where the deformation gradients are even smaller.

\begin{center}
\textbf{\uppercase{\small S-II. Numerical scheme}}
\end{center}

We discretized Eq. 16 of the main text using the standard second-order finite difference formulae and the time integration is performed using the semi-implicit Crank-Nicolson (C-N) method \cite{LeVeque2007SM}. We use uniform grid spacing $\Delta x$ and $\Delta t$ for the space and time, respectively. Applying the C-N method to Eq. 16 of the main text we obtain 

\begin{eqnarray}
\label{eq:CN_discretization}
& & -(\mathcal{D}_j + p_2 \mathcal{U}_j) r_{j+1}^{m+1} + (p_1 + 2\mathcal{D}_j) r_{j}^{m+1} - (\mathcal{D}_j - p_2 \mathcal{U}_j) r_{j-1}^{m+1} \nonumber \\ 
& & ~~~~~~~~~~~~~~~~~~~~~~~~~~~~~~~~~~~~ = (\mathcal{D}_j + p_2 \mathcal{U}_j) r_{j+1}^{m} + (p_1 - 2\mathcal{D}_j) r_{j}^{m} + (\mathcal{D}_j - p_2 \mathcal{U}_j) r_{j-1}^{m} + p_3 \mathcal{F}_j,
\end{eqnarray}
where $p_1 = \displaystyle \frac{2 \Delta x^2}{\Delta t}, ~ p_2 = \frac{\Delta x}{2}, ~ p_3 = 2 \Delta x^2, ~ \mathcal{D}_j = \mathcal{D}(x_j), ~ \mathcal{U}_j = \mathcal{U}(x_j), ~ \mathcal{F}_j = \mathcal{F}(x_j), ~ r_j^{m} = r(x_j, m\Delta t)$. Eq. \eqref{eq:CN_discretization} is a tridiagonal system of the form, 
\begin{equation}
\label{eq:tridiagonal}
\mathbf{L} \mathbf{r}^{m+1} = \mathbf{R}, 
\end{equation}
where the entries of the tridiagonal matrices, $\mathbf{L}$ and $\mathbf{R}$, are 

\begin{subequations}
\begin{align}
& \mathbf{L}_{j,j} = (p_1 + 2\mathcal{D}_j), ~ \mathbf{L}_{j,j+1} = -(\mathcal{D}_j + p_2 \mathcal{U}_j), ~ \mathbf{L}_{j,j-1} = -(\mathcal{D}_j - p_2 \mathcal{U}_j), \label{eq:L_entries} \\
& \mathbf{R}_{j,j} = (p_1 - 2\mathcal{D}_j)r_j^m + p_3\mathcal{F}_j, ~ \mathbf{R}_{j,j+1} = (\mathcal{D}_j + p_2 \mathcal{U}_j)r_{j+1}^m, ~ \mathbf{R}_{j,j-1} = (\mathcal{D}_j - p_2 \mathcal{U}_j)r_{j-1}^m, \label{eq:R_entries}
\end{align}
\end{subequations}
\end{widetext}
and $\mathbf{r}^{m+1}$ is the column vector of the unknown variables ($r_j$) at the $(m+1)$-th time step. A (local) von Neumann stability analysis of the corresponding homogeneous equation of Eq. \eqref{eq:CN_discretization} (i.e. $\mathcal{F}_j = 0, ~ \forall j$) gives \cite{LeVeque2007SM, Toppaladoddi2017SM} 

\begin{subequations}
\begin{align}
& \lvert G_j \rvert^2 = \frac{\left(a_{+,j} a_{-,j} - b_j^2\right)^2 + 4p_1^2b_j^2}{\left(a_{+,j}^2 + b_j^2\right)^2}, \label{eq:Gsquare} \\
& a_{\pm,j} = p_1 \pm 2 D_j \sin^2\left(\frac{\Delta x}{2}\right), ~~~ b_j = 2p_2\mathcal{U}_j. \label{eq:aj_bj}
\end{align}
\end{subequations}

\begin{enumerate}
\item When $b_j \ll 1$ (in other words, $\mathcal{U}_j \ll 1$), we have 

\begin{equation}
\lvert G_j \rvert^2 \approx \left(\frac{a_{-,j}}{a_{+,j}}\right)^2 = \left[\frac{p_1 - 2 D_j \sin^2\left(\frac{\Delta x}{2}\right)}{p_1 + 2 D_j \sin^2\left(\frac{\Delta x}{2}\right)}\right]^2. 
\end{equation}
This is the amplification factor of the pure diffusion equation for C-N method, and implies an unconditional stability for all values of $\Delta x$ and $\Delta t$. 

\item When $b_j \gg 1$ (in other words, $\mathcal{U}_j \gg 1$), we have 

\begin{equation}
\lvert G_j \rvert^2 \approx 1 + \frac{4p_1^2}{b_{j}^2}, 
\end{equation}
i.e. the C-N scheme is numerically unstable. 
\end{enumerate} 

The results from the numerical computations presented in \S IV B of the main text were obtained using (a) $n = 3$, $\eta_r = 10^2, 10^3, 10^4, 10^5, 10^6, 10^7$, $d_0 = 8$ \AA, $\lambda_3 = 5.3947$ \AA, (b) $n = 2$, $\eta_r = 10^7$, $d_0 = 8$ \AA, $\lambda_2 = 5.3947$ \AA, a temperature gradient $\mathcal{G} = 1$ K/m (or $10^2$ K/m) and final (dimensional) time of integration $T_f = 10^2$ h. 

The numerical solutions with the homogeneous Dirichlet boundary condition 

\begin{equation}
\label{eq:DBC_x1}
r = 0, ~~ x = 1 
\end{equation}
causes a boundary layer formation near the cold end at $x = 1$. Numerical computations are also performed using 

\begin{equation}
\label{eq:relaxed_BC}
\partial_{xx} r = 0, ~~ x = 1, 
\end{equation}
in place of Eq. \eqref{eq:DBC_x1}. With the latter boundary condition a boundary layer does not form. However, it is observed that the capillary deformation near the bulk solid-liquid interface remains unaffected due to the choice of the boundary condition at $x = 1$. The results shown in \S IV B are obtained using \eqref{eq:DBC_x1}. 

\begin{center}
\textbf{\uppercase{\small S-III. Parameters}}
\end{center}

\begin{center}
\textbf{\small S-A. Lists of parameters used in the numerical solutions for the ice-water system}
\end{center}

\begin{enumerate}

\item $\lambda_3 = 5.3947$ \AA, $d_0 = 1$ \AA, $\eta_r = 10^2$. 

\begin{center}
\begin{tabular}{|c|c|c|}
\hline 
$\gamma$ & $\left(r_2 - r_0\right)_{\rm max}(T_f)$ [$\mu$m] & $x_{\rm max}$ [$\mu$m] \\
\hline
0.1 & 0.0805 & $0.4500 \times 10^3$ \\ 
0.25 & 0.1108 & $0.6000 \times 10^3$ \\ 
0.5 & 0.1774 & $0.9250 \times 10^3$ \\ 
0.75 & 0.2470 & $1.3000 \times 10^3$ \\ 
1 & 0.2863 & $1.5500 \times 10^3$ \\ 
1.25 & 0.2970 & $1.6500 \times 10^3$ \\ 
1.5 & 0.2987 & $1.6750 \times 10^3$ \\ 
\hline  
\end{tabular}
\end{center}

\item $\lambda_3 = 5.3947$ \AA, $d_0 = 8$ \AA, $\eta_r = 10^2$. 

\begin{center}
\begin{tabular}{|c|c|c|}
\hline 
$\gamma$ & $\left(r_2 - r_0\right)_{\rm max}(T_f)$ [$\mu$m] & $x_{\rm max}$ [$\mu$m] \\
\hline
0.1 & 0.0753 & $0.4200 \times 10^3$ \\ 
0.25 & 0.0943 & $0.5100 \times 10^3$ \\ 
0.5 & 0.1334 & $0.7000 \times 10^3$ \\ 
0.75 & 0.1804 & $0.9250 \times 10^3$ \\ 
1 & 0.2289 & $1.1750 \times 10^3$ \\ 
1.5 & 0.2902 & $1.5500 \times 10^3$ \\ 
1.75 & 0.2984 & $1.6250 \times 10^3$ \\ 
1.8 & 0.2991 & $1.6250 \times 10^3$ \\ 
1.9 & 0.2998 & $1.6500 \times 10^3$ \\ 
2 & 0.3001 & $1.6500 \times 10^3$ \\ 
2.5 & 0.2994 & $1.6750 \times 10^3$ \\ 
2.75 & 0.2991 & $1.6750 \times 10^3$ \\ 
2.275 & 0.2998 & $1.6750 \times 10^3$ \\ 
\hline  
\end{tabular}
\end{center}

\item $\lambda_3 = 5.3947$ \AA, $d_0 = 1$ \AA, $\eta_r = 10^3$. 

\begin{center}
\begin{tabular}{|c|c|c|}
\hline 
$\gamma$ & $\left(r_2 - r_0\right)_{\rm max}(T_f)$ [$\mu$m] & $x_{\rm max}$ [$\mu$m] \\
\hline
0.1 & 0.0378 & $0.2100 \times 10^3$ \\ 
0.2 & 0.0475 & $0.2600 \times 10^3$ \\ 
0.25 & 0.0532 & $0.2850 \times 10^3$ \\ 
0.33 & 0.0639 & $0.3400 \times 10^3$ \\ 
0.5 & 0.0910 & $0.4750 \times 10^3$ \\ 
0.67 & 0.1268 & $0.6500 \times 10^3$ \\ 
0.75 & 0.1481 & $0.7500 \times 10^3$ \\ 
1 & 0.2190 & $1.1000 \times 10^3$ \\ 
1.5 & 0.2960 & $1.6250 \times 10^3$ \\ 
2 & 0.2991 & $1.6750 \times 10^3$ \\ 
\hline  
\end{tabular}
\end{center}

\item $\lambda_3 = 5.3947$ \AA, $d_0 = 8$ \AA, $\eta_r = 10^3$. 
\begin{center}
\begin{tabular}{|c|c|c|}
\hline 
$\gamma$ & $\left(r_2 - r_0\right)_{\rm max}(T_f)$ [$\mu$m] & $x_{\rm max}$ [$\mu$m] \\
\hline
0.1 & 0.0353 & $0.1950 \times 10^3$ \\ 
0.2 & 0.0415 & $0.2250 \times 10^3$ \\ 
0.25 & 0.0449 & $0.2450 \times 10^3$ \\ 
0.33 & 0.0512 & $0.2750 \times 10^3$ \\ 
0.5 & 0.0659 & $0.3400 \times 10^3$ \\ 
0.67 & 0.0839 & $0.4300 \times 10^3$ \\ 
0.75 & 0.0942 & $0.4700 \times 10^3$ \\ 
1 & 0.1308 & $0.6400 \times 10^3$ \\ 
1.5 & 0.2232 & $1.0750 \times 10^3$ \\ 
2 & 0.2921 & $1.5000 \times 10^3$ \\ 
2.1 & 0.2979 & $1.5500 \times 10^3$ \\ 
2.15 & 0.2998 & $1.5750 \times 10^3$ \\ 
2.25 & 0.3023 & $1.6000 \times 10^3$ \\ 
2.5 & 0.3033 & $1.6500 \times 10^3$ \\ 
2.7 & 0.3021 & $1.6750 \times 10^3$ \\ 
3 & 0.3004 & $1.6750 \times 10^3$ \\ 
3.25 & 0.2996 & $1.6750 \times 10^3$ \\ 
3.5 & 0.2902 & $1.6750 \times 10^3$ \\ 
\hline  
\end{tabular}
\end{center}

\item $\lambda_3 = 5.3947$ \AA, $d_0 = 1$ \AA, $\eta_r = 10^4$. 

\begin{center}
\begin{tabular}{|c|c|c|}
\hline 
$\gamma$ & $\left(r_2 - r_0\right)_{\rm max}(T_f)$ [$\mu$m] & $x_{\rm max}$ [$\mu$m] \\
\hline
0.25 & 0.0252 & $0.1350 \times 10^3$ \\ 
0.33 & 0.0306 & $0.1625 \times 10^3$ \\ 
0.5 & 0.0443 & $0.2300 \times 10^3$ \\ 
0.67 & 0.0633 & $0.3225 \times 10^3$ \\
0.75 & 0.0753 & $0.3800 \times 10^3$ \\ 
1 & 0.1228 & $0.6000 \times 10^3$ \\ 
1.5 & 0.2558 & $1.2750 \times 10^3$ \\ 
2 & 0.3004 & $1.6500 \times 10^3$ \\ 
2.5 & 0.2992 & $1.6750 \times 10^3$ \\ 
3 & 0.2988 & $1.6750 \times 10^3$ \\ 
\hline 
\end{tabular}
\end{center}

\item $\lambda_3 = 5.3947$ \AA, $d_0 = 8$ \AA, $\eta_r = 10^4$. 

\begin{center}
\begin{tabular}{|c|c|c|}
\hline 
$\gamma$ & $\left(r_2 - r_0\right)_{\rm max}(T_f)$ [$\mu$m] & $x_{\rm max}$ [$\mu$m] \\
\hline
0.25 & 0.0213 & $0.1150 \times 10^3$ \\ 
0.33 & 0.0245 & $0.1300 \times 10^3$ \\ 
0.5 & 0.0319 & $0.1650 \times 10^3$ \\ 
0.67 & 0.0413 & $0.2100 \times 10^3$ \\ 
0.75 & 0.0467 & $0.2350 \times 10^3$ \\ 
1 & 0.0669 & $0.3275 \times 10^3$ \\ 
1.5 & 0.1280 & $0.5950 \times 10^3$ \\ 
2 & 0.2166 & $1.000 \times 10^3$ \\ 
2.5 & 0.2931 & $1.4250 \times 10^3$ \\ 
2.55 & 0.2974 & $1.4500 \times 10^3$ \\ 
2.6 & 0.3010 & $1.5000 \times 10^3$ \\ 
2.75 & 0.3075 & $1.5750 \times 10^3$ \\ 
3 & 0.3087 & $1.6500 \times 10^3$ \\ 
3.25 & 0.3053 & $1.6750 \times 10^3$ \\ 
3.45 & 0.3028 & $1.6750 \times 10^3$ \\ 
4 & 0.2996 & $1.6500 \times 10^3$ \\ 
4.25 & 0.2992 & $1.6750 \times 10^3$ \\ 
4.5 & 0.2990 & $1.6750 \times 10^3$ \\ 
\hline 
\end{tabular}
\end{center}

\item $\lambda_3 = 5.3947$ \AA, $d_0 = 8$ \AA, $\eta_r = 10^5$. 

\begin{center}
\begin{tabular}{|c|c|c|}
\hline 
$\gamma$ & $\left(r_2 - r_0\right)_{\rm max}(T_f)$ [$\mu$m] & $x_{\rm max}$ [$\mu$m] \\
\hline 
2.75 & 0.2546 & $1.1250 \times 10^3$ \\ 
3 & 0.2918 & $1.3500 \times 10^3$ \\ 
3.05 & 0.2974 & $1.4000 \times 10^3$ \\
3.1 & 0.3023 & $1.4250 \times 10^3$ \\ 
3.75 & 0.3113 & $1.6750 \times 10^3$ \\ 
4 & 0.3058 & $1.7000 \times 10^3$ \\
4.15 & 0.3034 & $1.7000 \times 10^3$ \\
4.75 & 0.2995 & $1.6750 \times 10^3$ \\ 
5 & 0.2991 & $1.6750 \times 10^3$ \\
5.25 & 0.2989 & $1.6750 \times 10^3$ \\
\hline 
\end{tabular}
\end{center}

\item $\lambda_3 = 5.3947$ \AA, $d_0 = 8$ \AA, $\eta_r = 10^6$. 

\begin{center}
\begin{tabular}{|c|c|c|}
\hline 
$\gamma$ & $\left(r_2 - r_0\right)_{\rm max}(T_f)$ [$\mu$m] & $x_{\rm max}$ [$\mu$m] \\
\hline 
3.25 & 0.2466 & $1.0500 \times 10^3$ \\ 
3.55 & 0.2943 & $1.3000 \times 10^3$ \\ 
3.575 & 0.2975 & $1.3250 \times 10^3$ \\ 
3.625 & 0.3034 & $1.3750 \times 10^3$ \\ 
4 & 0.3241 & $1.6000 \times 10^3$ \\ 
4.25 & 0.3198 & $1.6750 \times 10^3$ \\ 
4.5 & 0.3116 & $1.7000 \times 10^3$ \\ 
4.8 & 0.3046 & $1.7000 \times 10^3$ \\ 
5 & 0.3020 & $1.7000 \times 10^3$ \\ 
5.5 & 0.2994 & $1.6750 \times 10^3$ \\ 
5.75 & 0.2991 & $1.6750 \times 10^3$ \\ 
6 & 0.2989 & $1.6750 \times 10^3$ \\ 
\hline 
\end{tabular}
\end{center}

\item $\lambda_3 = 5.3947$ \AA, $d_0 = 1$ \AA, $\eta_r = 10^7$. 

\begin{center}
\begin{tabular}{|c|c|c|}
\hline 
$\gamma$ & $\left(r_2 - r_0\right)_{\rm max}(T_f)$ [$\mu$m] & $x_{\rm max}$ [$\mu$m] \\
\hline 
0.25 & 0.0027 & $0.1450 \times 10^2$ \\ 
0.33 & 0.0033 & $0.1750 \times 10^2$ \\ 
0.5 & 0.0050 & $0.2600 \times 10^2$ \\ 
0.67 & 0.0074 & $0.3800 \times 10^2$ \\ 
0.75 & 0.0090 & $0.4550 \times 10^2$ \\
1 & 0.0158 & $0.7700 \times 10^2$ \\ 
1.5 & 0.0442 & $2.0500 \times 10^2$ \\ 
2 & 0.1111 & $4.9500 \times 10^2$ \\ 
2.5 & 0.2350 & $1.0500 \times 10^3$ \\ 
2.75 & 0.2919 & $1.3750 \times 10^3$ \\ 
3 & 0.3109 & $1.6000 \times 10^3$ \\ 
4 & 0.2990 & $1.67500 \times 10^3$ \\ 
\hline 
\end{tabular}
\end{center}

\item $\lambda_2 = 5.3947$ \AA, $d_0 = 8$ \AA, $\eta_r = 10^7$. 

\begin{center}
\begin{tabular}{|c|c|c|}
\hline 
$\gamma$ & $\left(r_2 - r_0\right)_{\rm max}(T_f)$ [$\mu$m] & $x_{\rm max}$ [$\mu$m] \\
\hline 
0.05 & 0.0239 & $0.110 \times 10^3$ \\ 
0.5 & 0.0598 & $0.250 \times 10^3$ \\ 
1 & 0.1456 & $0.600 \times 10^3$ \\ 
1.5 & 0.3175 & $1.250 \times 10^3$ \\ 
2 & 0.6314 & $2.425 \times 10^3$ \\ 
2.1 & 0.7173 & $2.700 \times 10^3$ \\ 
2.25 & 0.8623 & $3.300 \times 10^3$ \\ 
2.5 & 1.1500 & $4.300 \times 10^3$ \\ 
2.75 & 1.4867 & $5.600 \times 10^3$ \\ 
3 & 1.8372 & $6.975 \times 10^3$ \\ 
3.25 & 2.1221 & $8.400 \times 10^3$ \\ 
3.5 & 2.2547 & $9.400 \times 10^3$ \\ 
3.75 & 2.2459 & $9.900 \times 10^3$ \\ 
4 & 2.1975 & $1.000 \times 10^4$ \\ 
4.25 & 2.1667 & $1.000 \times 10^4$ \\ 
4.5 & 2.1535 & $9.900 \times 10^3$ \\ 
4.75 & 2.1486 & $9.900 \times 10^3$ \\
5 & 2.1469 & $9.900 \times 10^3$ \\ 
\hline 
\end{tabular}
\end{center}

\item $\lambda_2 = 0.2101$ \AA, $d_0 = 1$ \AA, $\eta_r = 10^2$. 

\begin{center}
\begin{tabular}{|c|c|c|}
\hline 
$\gamma$ & $\left(r_2 - r_0\right)_{\rm max}(T_f)$ [$\mu$m] & $x_{\rm max}$ [$\mu$m] \\
\hline 
$(1, ~ 0)$ & 0.1329 & $0.6100 \times 10^3$ \\ 
1/4 & 0.0545 & $0.2450 \times 10^3$ \\ 
1/2 & 0.0788 & $0.3450 \times 10^3$ \\ 
1 & 0.1232 & $0.5400 \times 10^3$ \\ 
3/2 & 0.1327 & $0.6100 \times 10^3$ \\ 
\hline
\end{tabular}
\end{center}

\item $\lambda_2 = 0.2101$ \AA, $d_0 = 3$ \AA, $\eta_r = 10^2$. 

\begin{center}
\begin{tabular}{|c|c|c|}
\hline 
$\gamma$ & $\left(r_2 - r_0\right)_{\rm max}(T_f)$ [$\mu$m] & $x_{\rm max}$ [$\mu$m] \\
\hline 
1/2 & 0.0693 & $0.3000 \times 10^3$ \\ 
1 & 0.1100 & $0.4700 \times 10^3$ \\ 
3/2 & 0.1309 & $0.5800 \times 10^3$ \\ 
2 & 0.1332 & $0.6100 \times 10^3$ \\ 
5/2 & 0.1330 & $0.6100 \times 10^3$ \\ 
\hline
\end{tabular}
\end{center}

\item $\lambda_2 = 0.2101$ \AA, $d_0 = 5$ \AA, $\eta_r = 10^2$. 

\begin{center}
\begin{tabular}{|c|c|c|}
\hline 
$\gamma$ & $\left(r_2 - r_0\right)_{\rm max}(T_f)$ [$\mu$m] & $x_{\rm max}$ [$\mu$m] \\
\hline 
1/2 & 0.0651 & $0.2850 \times 10^3$ \\ 
1 & 0.1018 & $0.4300 \times 10^3$ \\ 
3/2 & 0.1274 & $0.5600 \times 10^3$ \\ 
2 & 0.1335 & $0.60500 \times 10^3$ \\ 
5/2 & 0.1334 & $0.6150 \times 10^3$ \\ 
\hline
\end{tabular}
\end{center} 

\item $\lambda_2 = 0.2101$ \AA, $d_0 = 8$ \AA, $\eta_r = 10^2$. 

\begin{center}
\begin{tabular}{|c|c|c|}
\hline 
$\gamma$ & $\left(r_2 - r_0\right)_{\rm max}(T_f)$ [$\mu$m] & $x_{\rm max}$ [$\mu$m] \\
\hline 
1/2 & 0.0614 & $0.2675 \times 10^3$ \\
1/2 & 0.0614 & $0.2650 \times 10^3$ \\ 
1 & 0.0936 & $0.3950 \times 10^3$ \\ 
3/2 & 0.1213 & $0.5200 \times 10^3$ \\ 
2 & 0.1330 & $0.5900 \times 10^3$ \\ 
5/2 & 0.1342 & $0.6100 \times 10^3$ \\
3 & 0.1335 & $0.6100 \times 10^3$ \\ 
\hline
\end{tabular}
\end{center}

\item $\lambda_2 = 0.2101$ \AA, $d_0 = 10$ \AA, $\eta_r = 10^2$. 

\begin{center}
\begin{tabular}{|c|c|c|}
\hline 
$\gamma$ & $\left(r_2 - r_0\right)_{\rm max}(T_f)$ [$\mu$m] & $x_{\rm max}$ [$\mu$m] \\
\hline 
1/2 & 0.0597 & $0.2600 \times 10^3$ \\ 
1 & 0.0896 & $0.3750 \times 10^3$ \\ 
3/2 & 0.1173 & $0.4900 \times 10^3$ \\ 
2 & 0.1320 & $0.5700 \times 10^3$ \\ 
5/2 & 0.1347 & $0.6100 \times 10^3$ \\ 
3 & 0.1340 & $0.6100 \times 10^3$ \\ 
\hline
\end{tabular}
\end{center}

\item $\lambda_{3/2} = 0.0337$ \AA, $d_0 = 1$ \AA, $\eta_r = 10^2$. 

\begin{center}
\begin{tabular}{|c|c|c|}
\hline 
$\gamma$ & $\left(r_2 - r_0\right)_{\rm max}(T_f)$ [$\mu$m] & $x_{\rm max}$ [$\mu$m] \\
\hline 
$(1, ~ 0)$ & 0.2007 & $0.8300 \times 10^3$ \\ 
1/4 & 0.0931 & $0.3750 \times 10^3$ \\ 
1/2 & 0.1294 & $0.5100 \times 10^3$ \\ 
1 & 0.1895 & $0.7500 \times 10^3$ \\ 
3/2 & 0.2006 & $0.8200 \times 10^3$ \\ 
\hline
\end{tabular}
\end{center}

\item $\lambda_{3/2} = 0.0337$ \AA, $d_0 = 3$ \AA, $\eta_r = 10^2$. 

\begin{center}
\begin{tabular}{|c|c|c|}
\hline 
$\gamma$ & $\left(r_2 - r_0\right)_{\rm max}(T_f)$ [$\mu$m] & $x_{\rm max}$ [$\mu$m] \\
\hline 
1/2 & 0.1159 & $0.4550 \times 10^3$ \\ 
1/2 & 0.1159 & $0.4600 \times 10^3$ \\ 
1 & 0.1733 & $0.6700 \times 10^3$ \\ 
3/2 & 0.1988 & $0.8000 \times 10^3$ \\ 
2 & 0.2011 & $0.8300 \times 10^3$ \\ 
\hline
\end{tabular}
\end{center}

\item $\lambda_{3/2} = 0.0337$ \AA, $d_0 = 5$ \AA, $\eta_r = 10^2$. 

\begin{center}
\begin{tabular}{|c|c|c|}
\hline 
$\gamma$ & $\left(r_2 - r_0\right)_{\rm max}(T_f)$ [$\mu$m] & $x_{\rm max}$ [$\mu$m] \\
\hline 
1/2 & 0.1098 & $0.4300 \times 10^3$ \\ 
1 & 0.1627 & $0.6300 \times 10^3$ \\ 
3/2 & 0.1952 & $0.7750 \times 10^3$ \\ 
2 & 0.2015 & $0.8200 \times 10^3$ \\ 
5/2 & 0.2012 & $0.8300 \times 10^3$ \\ 
\hline
\end{tabular}
\end{center}

\item $\lambda_{3/2} = 0.0337$ \AA, $d_0 = 8$ \AA, $\eta_r = 10^2$. 

\begin{center}
\begin{tabular}{|c|c|c|}
\hline 
$\gamma$ & $\left(r_2 - r_0\right)_{\rm max}(T_f)$ [$\mu$m] & $x_{\rm max}$ [$\mu$m] \\
\hline 
1/2 & 0.1044 & $0.4100 \times 10^3$ \\ 
1 & 0.1519 & $0.5800 \times 10^3$ \\ 
3/2 & 0.1884 & $0.7300 \times 10^3$ \\ 
2 & 0.2014 & $0.8100 \times 10^3$ \\ 
5/2 & 0.2021 & $0.8300 \times 10^3$ \\ 
3 & 0.2013 & $0.8300 \times 10^3$ \\ 
\hline
\end{tabular}
\end{center}

\item $\lambda_{3/2} = 0.0337$ \AA, $d_0 = 10$ \AA, $\eta_r = 10^2$. 

\begin{center}
\begin{tabular}{|c|c|c|}
\hline 
$\gamma$ & $\left(r_2 - r_0\right)_{\rm max}(T_f)$ [$\mu$m] & $x_{\rm max}$ [$\mu$m] \\
\hline 
1/2 & 0.1020 & $0.4000 \times 10^3$ \\ 
1 & 0.1465 & $0.5600 \times 10^3$ \\ 
3/2 & 0.1838 & $0.7000 \times 10^3$ \\ 
2 & 0.2006 & $0.7900 \times 10^3$ \\ 
5/2 & 0.2027 & $0.8200 \times 10^3$ \\ 
3 & 0.2018 & $0.8300 \times 10^3$ \\ 
\hline
\end{tabular}
\end{center}

\item $\lambda_{3/2} = 1.3759$ \AA, $d_0 = 1$ \AA, $\eta_r = 10^2$. 

\begin{center}
\begin{tabular}{|c|c|c|}
\hline 
$\gamma$ & $\left(r_2 - r_0\right)_{\rm max}(T_f)$ [$\mu$m] & $x_{\rm max}$ [$\mu$m] \\
\hline 
$(1, ~ 0)$ & 0.0762 & $0.4300 \times 10^3$ \\ 
1/4 & 0.0263 & $0.1425 \times 10^3$ \\
1/2 & 0.0401 & $0.2100 \times 10^3$ \\ 
1 & 0.0687 & $0.3600 \times 10^3$ \\ 
3/2 & 0.0760 & $0.4200 \times 10^3$ \\ 
2 & 0.0762 & $0.4300 \times 10^3$ \\ 
\hline
\end{tabular}
\end{center}

\item $\lambda_{3} = 1.3759$ \AA, $d_0 = 3$ \AA, $\eta_r = 10^2$. \\ 

\begin{center}
\begin{tabular}{|c|c|c|}
\hline 
$\gamma$ & $\left(r_2 - r_0\right)_{\rm max}(T_f)$ [$\mu$m] & $x_{\rm max}$ [$\mu$m] \\
\hline 
1/2 & 0.0344 & $0.1800 \times 10^3$ \\ 
1 & 0.0591 & $0.3000 \times 10^3$ \\ 
3/2 & 0.0743 & $0.4000 \times 10^3$ \\ 
2 & 0.0765 & $0.4200 \times 10^3$ \\ 
\hline
\end{tabular}
\end{center}

\item $\lambda_{3} = 1.3759$ \AA, $d_0 = 5$ \AA, $\eta_r = 10^2$. 

\begin{center}
\begin{tabular}{|c|c|c|}
\hline 
$\gamma$ & $\left(r_2 - r_0\right)_{\rm max}(T_f)$ [$\mu$m] & $x_{\rm max}$ [$\mu$m] \\
\hline 
1/2 & 0.0319 & $0.1650 \times 10^3$ \\ 
1 & 0.0535 & $0.2700 \times 10^3$ \\ 
3/2 & 0.0713 & $0.3700 \times 10^3$ \\ 
2 & 0.0766 & $0.4200 \times 10^3$ \\ 
5/2 & 0.0767 & $0.4300 \times 10^3$ \\ 
\hline
\end{tabular}
\end{center}

\item $\lambda_{3} = 1.3759$ \AA, $d_0 = 8$ \AA, $\eta_r = 10^2$. 

\begin{center}
\begin{tabular}{|c|c|c|}
\hline 
$\gamma$ & $\left(r_2 - r_0\right)_{\rm max}(T_f)$ [$\mu$m] & $x_{\rm max}$ [$\mu$m] \\
\hline 
1/2 & 0.0298 & $0.1550 \times 10^3$ \\ 
1 & 0.0482 & $0.2400 \times 10^3$ \\ 
3/2 & 0.0664 & $0.3300 \times 10^3$ \\ 
2 & 0.0759 & $0.3950 \times 10^3$ \\ 
5/2 & 0.0773 & $0.4200 \times 10^3$ \\ 
3 & 0.0768 & $0.4300 \times 10^3$ \\ 
4 & 0.0763 & $0.4300 \times 10^3$ \\ 
\hline
\end{tabular}
\end{center}

\item $\lambda_{3} = 1.3759$ \AA, $d_0 = 10$ \AA, $\eta_r = 10^2$. 

\begin{center}
\begin{tabular}{|c|c|c|}
\hline 
$\gamma$ & $\left(r_2 - r_0\right)_{\rm max}(T_f)$ [$\mu$m] & $x_{\rm max}$ [$\mu$m] \\
\hline 
1/2 & 0.0288 & $0.1500 \times 10^3$ \\ 
1 & 0.0456 & $0.2250 \times 10^3$ \\ 
3/2 & 0.0633 & $0.3100 \times 10^3$ \\ 
2 & 0.0747 & $0.3000 \times 10^3$ \\ 
5/2 & 0.0777 & $0.4150 \times 10^3$ \\ 
3 & 0.0773 & $0.4300 \times 10^3$ \\ 
\hline
\end{tabular}
\end{center}

\end{enumerate} 


\begin{center}
\textbf{\small S-B. Parameters for Dodecane}
\end{center}

\begin{subequations}
\begin{align}
& \mathcal{G} = 1 ~ {\rm K/m}, \label{eq:para1_D} \\
& k = 3 \times 10^9 ~ {\rm Pa/m}, \label{eq:para2_D} \\
& \rho_l = 0.78 \times 10^3 ~ {\rm kg/m}^{3}, \label{eq:para3_D} \\
& \rho_s = 0.78 \times 10^3 ~ {\rm kg/m}^3, \label{eq:para4_D} \\
& T_m = 263.2 - 263.8 ~ K, \label{eq:para5_D} \\
& \eta_b = 1.34 \times 10^{-3} ~ {\rm kg/ms}, \label{eq:para7_D} \\
& q_m = 216.2565 \times 10^3 ~ {\rm J/kg}. \label{eq:para8_D}
\end{align}
\end{subequations}

\begin{widetext}

\begin{table}[h]
\begin{tabular}{|c|c|c|c|c|c|c|c|c|c|c|c|}
\hline 
$n$ & $d_0$ [\AA] & $\lambda_n$ [\AA] & $d_0$ [\AA] & $\lambda_n$ [\AA] & $d_0$ [\AA] & $\lambda_n$ [\AA] & $d_0$ [\AA] & $\lambda_n$ [\AA] & $d_0$ [\AA] & $\lambda_n$ [\AA] \\
\hline 
3/2 & 1 & 1.5623 & 3 & 2.2532 & 5 & 2.6714 & 8 & 3.1245 & 10 & 3.3658 \\ 
2 & ~ & 1.9762 & ~ & 3.4229 & ~ & 4.4189 & ~ & 5.5895 & ~ & 6.2493 \\ 
3 & ~ & 1.9841 & ~ & 4.1271 & ~ & 5.8015 & ~ & 7.9364 & ~ & 9.2093 \\ 
\hline 
\end{tabular}
\caption{$\lambda_n$ for different $n$ and $d_0$. Other relevant parameters are listed in Eqs \eqref{eq:para1_D}-\eqref{eq:para8_D}.} 
\end{table}

\begin{table}[h]
\begin{tabular}{|c|c|c|c|c|c|c|c|c|c|c|c|c|}
\hline 
$d_0$ [\AA] & $\lambda_3$ [\AA] & $r_{\rm max}(T_{\rm f})$ [$\mu$m] & $x_{\rm max}$ [$\mu$m] & $\lambda_2$ [\AA] & $r_{\rm max}(T_{\rm f})$ [$\mu$m] & $x_{\rm max}$ [$\mu$m] & $\lambda_{3/2}$ [\AA] & $r_{\rm max}(T_{\rm f})$ [$\mu$m] & $x_{\rm max}$ [$\mu$m] \\
\hline 
1 & 1.9841 & 0.0597 & $0.5900 \times 10^3$ & 1.9762 & 0.4943 & $4.0000 \times 10^3$ & 1.5623 & 1.9452 & $1.4000 \times 10^4$ \\ 
3 & 4.1271 & 0.1243 & $1.2250 \times 10^3$ & 3.4229 & 0.7914 & $6.4000 \times 10^3$ & 2.2532 & 2.5599 & $1.8500 \times 10^4$ \\ 
5 & 5.8015 & 0.1747 & $1.7000 \times 10^3$ & 4.4189 & 0.9851 & $8.0000 \times 10^3$ & 2.6714 & 2.9086 & $2.1000 \times 10^4$ \\ 
8 & 7.9364 & 0.2390 & $2.3500 \times 10^3$ & 5.5895 & 1.2051 & $9.7500 \times 10^3$ & 3.1245 & 3.2711 & $2.3500 \times 10^4$ \\ 
10 & 9.2093 & 0.2773 & $2.7500 \times 10^3$ & 6.2493 & 1.3260 & $1.0750 \times 10^4$ & 3.3658 & 3.4589 & $2.5000 \times 10^4$ \\
\hline 
\end{tabular}
\caption{$T_{\rm f} = 10^2$ h, $\eta = \eta_b$.} 
\end{table}

\end{widetext}


\begin{center}
\textit{S1. Lists of parameters used in the numerical solutions}
\end{center}

\begin{enumerate}

\item $\lambda_3 = 1.9841$, \AA, $d_0 = 1$ \AA, $\eta_r = 10^2$. 

\begin{center}
\begin{tabular}{|c|c|c|}
\hline 
$\gamma$ & $\left(r_2 - r_0\right)_{\rm max}(T_f)$ [$\mu$m] & $x_{\rm max}$ [$\mu$m] \\
\hline 
1/4 & 0.0211 & $0.2000 \times 10^3$ \\
1/2 & 0.0325 & $0.3000 \times 10^3$ \\
1 &  0.0550 & $0.5100 \times 10^3$ \\
3/2 & 0.0597 & $0.5800 \times 10^3$ \\
2 & 0.0598 & $0.5900 \times 10^3$ \\
\hline
\end{tabular}
\end{center}

\item $\lambda_3 = 4.1271$, \AA, $d_0 = 3$ \AA, $\eta_r = 10^2$. 

\begin{center}
\begin{tabular}{|c|c|c|}
\hline 
$\gamma$ & $\left(r_2 - r_0\right)_{\rm max}(T_f)$ [$\mu$m] & $x_{\rm max}$ [$\mu$m] \\
\hline 
1/4 & 0.0418 & $0.4000 \times 10^3$ \\ 
1/2 & 0.0622 & $0.5750 \times 10^3$ \\
1 & 0.1077 & $0.9750 \times 10^3$ \\ 
3/2 & 0.1236 & $1.2000 \times 10^3$ \\ 
2 & 0.1244 & $1.2250 \times 10^3$ \\ 
\hline 
\end{tabular}
\end{center}

\item $\lambda_3 = 5.8015$, \AA, $d_0 = 5$ \AA, $\eta_r = 10^2$. 

\begin{center}
\begin{tabular}{|c|c|c|}
\hline 
$\gamma$ & $\left(r_2 - r_0\right)_{\rm max}(T_f)$ [$\mu$m] & $x_{\rm max}$ [$\mu$m] \\
\hline 
1/4 & 0.0574 & $0.5400 \times 10^3$ \\ 
1/2 & 0.0841 & $0.7750 \times 10^3$ \\ 
1 & 0.1458 & $1.3250 \times 10^3$ \\ 
3/2 & 0.1730 & $1.6500 \times 10^3$ \\ 
2 & 0.1750 & $1.7000 \times 10^3$ \\ 
\hline 
\end{tabular}
\end{center}

\item $\lambda_3 = 7.9364$, \AA, $d_0 = 8$ \AA, $\eta_r = 10^2$. 

\begin{center}
\begin{tabular}{|c|c|c|}
\hline 
$\gamma$ & $\left(r_2 - r_0\right)_{\rm max}(T_f)$ [$\mu$m] & $x_{\rm max}$ [$\mu$m] \\
\hline 
1/4 & 0.0770 & $0.7300 \times 10^3$ \\ 
1/2 & 0.1108 & $1.0250 \times 10^3$ \\ 
1 & 0.1916 & $1.7250 \times 10^3$ \\ 
3/2 & 0.2349 & $2.2000 \times 10^3$ \\ 
2 & 0.2397 & $2.3500 \times 10^3$ \\ 
\hline
\end{tabular}
\end{center}

\item $\lambda_3 = 9.2093$, \AA, $d_0 = 10$ \AA, $\eta_r = 10^2$. 

\begin{center}
\begin{tabular}{|c|c|c|}
\hline 
$\gamma$ & $\left(r_2 - r_0\right)_{\rm max}(T_f)$ [$\mu$m] & $x_{\rm max}$ [$\mu$m] \\
\hline 
1/4 & 0.0884 & $0.8250 \times 10^3$ \\ 
1/2 & 0.1263 & $1.1500 \times 10^3$ \\ 
1 & 0.2178 & $1.9500 \times 10^3$ \\ 
3/2 & 0.2713 & $2.5500 \times 10^3$ \\ 
2 & 0.2783 & $2.7000 \times 10^3$ \\ 
\hline 
\end{tabular}
\end{center}

\item $\lambda_{3/2} = 1.5623$ \AA, $d_0 = 1$ \AA, $\eta_r = 10^2$. 

\begin{center}
\begin{tabular}{|c|c|c|}
\hline 
$\gamma$ & $\left(r_2 - r_0\right)_{\rm max}(T_f)$ [$\mu$m] & $x_{\rm max}$ [$\mu$m] \\
\hline 
1/4 & 1.0065 & $0.1000 \times 10^4$ \\ 
1/2 & 1.4884 & $1.0250 \times 10^4$ \\ 
1 & 1.9294 & $1.3750 \times 10^4$ \\ 
3/2 & 1.9451 & $1.4000 \times 10^4$ \\ 
2 & 1.9452 & $1.4000 \times 10^4$ \\ 
\hline
\end{tabular}
\end{center}

\item $\lambda_{3/2} = 2.2532$ \AA, $d_0 = 3$ \AA, $\eta_r = 10^2$. 

\begin{center}
\begin{tabular}{|c|c|c|}
\hline 
$\gamma$ & $\left(r_2 - r_0\right)_{\rm max}(T_f)$ [$\mu$m] & $x_{\rm max}$ [$\mu$m] \\
\hline 
1/4 & 1.2581 & $0.8750 \times 10^4$ \\ 
1/2 & 1.8139 & $1.2500 \times 10^4$ \\ 
1 & 2.5069 & $1.7750 \times 10^4$ \\ 
3/2 & 2.5596 & $1.8500 \times 10^4$ \\ 
2 & 2.5600 & $1.8500 \times 10^4$ \\ 
\hline 
\end{tabular}
\end{center}

\item $\lambda_{3/2} = 2.6714$ \AA, $d_0 = 5$ \AA, $\eta_r = 10^2$. 

\begin{center}
\begin{tabular}{|c|c|c|}
\hline 
$\gamma$ & $\left(r_2 - r_0\right)_{\rm max}(T_f)$ [$\mu$m] & $x_{\rm max}$ [$\mu$m] \\
\hline 
1/4 & 1.3954 & $0.9750 \times 10^4$ \\ 
1/2 & 1.9828 & $1.3750 \times 10^4$ \\ 
1 & 2.8160 & $1.9750 \times 10^4$ \\ 
3/2 & 2.9078 & $2.1000 \times 10^4$ \\ 
2 & 2.9088 & $2.1000 \times 10^4$ \\ 
\hline 
\end{tabular}
\end{center}

\item $\lambda_{3/2} = 3.1245$ \AA, $d_0 = 8$ \AA, $\eta_r = 10^2$. 

\begin{center}
\begin{tabular}{|c|c|c|}
\hline 
$\gamma$ & $\left(r_2 - r_0\right)_{\rm max}(T_f)$ [$\mu$m] & $x_{\rm max}$ [$\mu$m] \\
\hline 
1/4 & 1.5347 & $1.0750 \times 10^4$ \\ 
1/2 & 2.1491 & $1.4750 \times 10^4$ \\ 
1 & 3.1189 & $2.1750 \times 10^4$ \\ 
3/2 & 3.2694 & $2.3500 \times 10^4$ \\ 
\hline
\end{tabular}
\end{center}

\item $\lambda_{3/2} = 3.3658$ \AA, $d_0 = 10$ \AA, $\eta_r = 10^2$. 

\begin{center}
\begin{tabular}{|c|c|c|}
\hline 
$\gamma$ & $\left(r_2 - r_0\right)_{\rm max}(T_f)$ [$\mu$m] & $x_{\rm max}$ [$\mu$m] \\
\hline 
1/4 & 1.6054 & $1.1300 \times 10^4$ \\ 
1/2 & 2.2319 & $1.5500 \times 10^4$ \\ 
1 & 3.2673 & $2.2750 \times 10^4$ \\ 
3/2 & 3.4560 & $2.4750 \times 10^4$ \\ 
2 & 3.4598 & $2.5000 \times 10^4$ \\ 
\hline 
\end{tabular}
\end{center}

\item $\lambda_2 = 1.9762$ \AA, $d_0 = 1$ \AA, $\eta_r = 10^2$. 

\begin{center}
\begin{tabular}{|c|c|c|}
\hline 
$\gamma$ & $\left(r_2 - r_0\right)_{\rm max}(T_f)$ [$\mu$m] & $x_{\rm max}$ [$\mu$m] \\
\hline 
1/4 & 0.2204 & $1.7250 \times 10^3$ \\ 
1/2 & 0.3369 & $2.6000 \times 10^3$ \\ 
1 & 0.4840 & $3.8000 \times 10^3$ \\ 
3/2 & 0.4942 & $4.0000 \times 10^3$ \\ 
2 & 0.4943 & $4.0000 \times 10^3$ \\ 
\hline
\end{tabular}
\end{center}

\item $\lambda_2 = 3.4229$, \AA, $d_0 = 3$ \AA, $\eta_r = 10^2$. 

\begin{center}
\begin{tabular}{|c|c|c|}
\hline 
$\gamma$ & $\left(r_2 - r_0\right)_{\rm max}(T_f)$ [$\mu$m] & $x_{\rm max}$ [$\mu$m] \\
\hline 
1/4 & 0.3353 & $2.6250 \times 10^3$ \\ 
1/2 & 0.4964 & $3.8000 \times 10^3$ \\ 
1 & 0.7555 & $5.9000 \times 10^3$ \\ 
3/2 & 0.7911 & $6.4000 \times 10^3$ \\ 
2 & 0.7916 & $6.4000 \times 10^3$ \\ 
\hline 
\end{tabular}
\end{center}

\item $\lambda_2 = 4.4189$, \AA, $d_0 = 5$ \AA, $\eta_r = 10^2$. 

\begin{center}
\begin{tabular}{|c|c|c|}
\hline 
$\gamma$ & $\left(r_2 - r_0\right)_{\rm max}(T_f)$ [$\mu$m] & $x_{\rm max}$ [$\mu$m] \\
\hline 
1/4 & 0.4074 & $3.2000 \times 10^3$ \\ 
1/2 & 0.5934 & $4.5000 \times 10^3$ \\ 
1 & 0.9220 & $7.2000 \times 10^3$ \\ 
3/2 & 0.9841 & $7.9000 \times 10^3$ \\ 
2 & 0.9854 & $8.0000 \times 10^3$ \\ 
\hline 
\end{tabular}
\end{center}

\item $\lambda_2 = 5.5895$, \AA, $d_0 = 8$ \AA, $\eta_r = 10^2$. 

\begin{center}
\begin{tabular}{|c|c|c|}
\hline 
$\gamma$ & $\left(r_2 - r_0\right)_{\rm max}(T_f)$ [$\mu$m] & $x_{\rm max}$ [$\mu$m] \\
\hline 
1/4 & 0.4874 & $3.8000 \times 10^3$ \\ 
1/2 & 0.6985 & $5.3000 \times 10^3$ \\ 
1 & 1.1009 & $8.5000 \times 10^3$ \\ 
3/2 & 1.2023 & $9.5000 \times 10^3$ \\ 
2 & 1.2057 & $9.7500 \times 10^3$ \\ 
\hline
\end{tabular}
\end{center}

\item $\lambda_3 = 1.9841$ \AA, $d_0 = 1$ \AA, $\eta_r = 10^4$. 

\begin{center}
\begin{tabular}{|c|c|c|}
\hline 
$\gamma$ & $\left(r_2 - r_0\right)_{\rm max}(T_f)$ [$\mu$m] & $x_{\rm max}$ [$\mu$m] \\
\hline 
(1,0) & 0.0598 & $0.6000 \times 10^3$ \\ 
3 & 0.0598 & $0.6000 \times 10^3$ \\ 
11/4 & 0.0599 & $0.6000 \times 10^3$ \\ 
5/2 & 0.0601 & $0.5750 \times 10^3$ \\ 
2 & 0.0595 & $0.5500 \times 10^3$ \\ 
7/4 & 0.0541 & $0.4750 \times 10^3$ \\ 
3/2 & 0.0430 & $0.3500 \times 10^3$ \\ 
1 & 0.0203 & $0.1750 \times 10^3$ \\ 
3/4 & 0.0129 &  $0.1150 \times 10^3$ \\ 
1/2 & 0.0080 & $0.0725 \times 10^3$ \\ 
\hline 
\end{tabular}
\end{center}

\item $\lambda_2 = 6.2493$, \AA, $d_0 = 10$ \AA, $\eta_r = 10^2$. 

\begin{center}
\begin{tabular}{|c|c|c|}
\hline 
$\gamma$ & $\left(r_2 - r_0\right)_{\rm max}(T_f)$ [$\mu$m] & $x_{\rm max}$ [$\mu$m] \\
\hline 
1/4 & 0.5307 & $4.2000 \times 10^3$ \\ 
1/2 & 0.7546 & $5.8000 \times 10^3$ \\ 
1 & 1.1947 & $9.2500 \times 10^3$ \\ 
3/2 & 1.3220 & $1.0500 \times 10^4$ \\ 
2 & 1.3270 & $1.0750 \times 10^4$ \\ 
\hline 
\end{tabular}
\end{center}

\item $\lambda_3 = 5.8015$ \AA, $d_0 = 5$ \AA, $\eta_r = 10^7$. 

\begin{center}
\begin{tabular}{|c|c|c|}
\hline 
$\gamma$ & $\left(r_2 - r_0\right)_{\rm max}(T_f)$ [$\mu$m] & $x_{\rm max}$ [$\mu$m] \\
\hline 
1/2 & 0.0023 & $0.2100 \times 10^2$ \\ 
3/4 & 0.0037 & $0.3250 \times 10^2$ \\ 
1 & 0.0058 & $0.4900 \times 10^2$ \\ 
3/2 & 0.0133 & $0.1075 \times 10^3$ \\ 
2 & 0.0281 & $0.2150 \times 10^3$ \\ 
5/2 & 0.0552 & $0.4100 \times 10^3$ \\ 
3 & 0.1002 & $0.7250 \times 10^3$ \\ 
3 & 0.1002 & $0.7300 \times 10^3$ \\ 
7/2 & 0.1580 & $1.1900 \times 10^3$ \\ 
4 & 0.1896 & $1.6250 \times 10^3$ \\ 
5 & 0.1761 & $1.7250 \times 10^3$ \\ 
6 & 0.1747 & $1.7250 \times 10^3$ \\ 
\hline 
\end{tabular}
\end{center}

\item $\lambda_3 = 5.8015$ \AA, $d_0 = 5$ \AA, $\eta_r = 10^4$. 

\begin{center}
\begin{tabular}{|c|c|c|}
\hline 
$\gamma$ & $\left(r_2 - r_0\right)_{\rm max}(T_f)$ [$\mu$m] & $x_{\rm max}$ [$\mu$m] \\
\hline 
(1,0) & 0.1747 & $1.7250 \times 10^3$ \\ 
3 & 0.1765 & $1.7250 \times 10^3$ \\ 
11/4 & 0.1781 & $1.7000 \times 10^3$ \\ 
5/2 & 0.1783 & $1.6500 \times 10^3$ \\ 
2 & 0.1518 & $1.2750 \times 10^3$ \\ 
7/4 & 0.1225 & $1.0000 \times 10^3$ \\ 
3/2 & 0.0920 & $0.7500 \times 10^3$ \\ 
1 & 0.0457 & $0.4000 \times 10^3$ \\ 
3/4 & 0.0308 &  $0.2700 \times 10^3$ \\ 
1/2 & 0.0203 & $0.1850 \times 10^3$ \\ 
1/3 & 0.0151 & $0.1400 \times 10^3$ \\ 
1/4 & 0.0130 & $0.1225 \times 10^3$ \\ 
1/5 & 0.0119 & $0.1125 \times 10^3$ \\ 
1/10 & 0.0098 & $0.0950 \times 10^3$ \\ 
\hline 
\end{tabular}
\end{center}

\end{enumerate}

\end{document}